\PassOptionsToPackage{dvipsnames}{xcolor}
\documentclass[sigconf, screen, authorversion]{acmart}

\setcopyright{acmlicensed}
\acmPrice{15.00}
\acmDOI{10.1145/3293882.3330576}
\acmYear{2019}
\copyrightyear{2019}
\acmISBN{978-1-4503-6224-5/19/07}
\acmConference[ISSTA '19]{Proceedings of the 28th ACM SIGSOFT International Symposium  on  Software Testing and Analysis}{July 15--19, 2019}{Beijing, China}
\acmBooktitle{Proceedings of the 28th ACM SIGSOFT International Symposium  on  Software Testing and Analysis (ISSTA '19), July 15--19, 2019, Beijing, China}

\usepackage[utf8]{inputenc}
\usepackage{framed}
\usepackage{url}
\usepackage{balance}
\usepackage{enumitem}
\usepackage{booktabs}
\usepackage{algorithm}
\usepackage{amsmath,amssymb,amsfonts,amsthm}
\usepackage{graphicx}
\usepackage{textcomp}
\usepackage{xcolor}
\usepackage{adjustbox}
\usepackage{subcaption}
\usepackage[noend]{algpseudocode}
\usepackage{listings}
\usepackage{tikz}
\usetikzlibrary{decorations.pathmorphing}
\usepackage{pifont}
\usepackage{multirow}

\newcommand{\tool}{{Zest}}

\newcommand{\code}{\texttt}
\newcommand{\cgf}{\text{CGF}}
\newcommand{\Q}{\mathcal{S}}

\newif\ifcomment
\commenttrue %
\ifcomment
    \newcommand{\nbc}[3]{
     {\colorbox{#3}{\bfseries\sffamily\scriptsize\textcolor{white}{#1}}}
     {\textcolor{#3}{\sf\small$\blacktriangleright$\textit{#2}$\blacktriangleleft$}}
     }
\else 
    \newcommand{\nbc}[3]{}
\fi

\ifcomment
    \newcommand\todo[1]{
    {\colorbox{red}{\bfseries\sffamily\scriptsize\textcolor{white}{TODO}}}
     {\textcolor{red}{\sf\small\textit{#1}}}
    }
\else
    \newcommand\todo[1]{}
\fi

\definecolor{clcolor}{rgb}{0.5,0.7,0.9}
\definecolor{rpcolor}{rgb}{0.05,0.4,0.1}
\definecolor{kscolor}{rgb}{0.9,0.1,0.1}
\definecolor{mcolor}{rgb}{1.0, 0.55, 0.0}

\newcommand\EPS[1]{\textcolor{Bittersweet}{\underline{\textbf{#1}}}}

\lstdefinelanguage{customC}{
  belowcaptionskip=1\baselineskip,
  breaklines=true,
  xleftmargin=\parindent,
  language=C,
  showstringspaces=false,
  basicstyle=\scriptsize\ttfamily,
  keywordstyle=\bfseries\color{green!40!black},
  commentstyle=\itshape\color{purple!40!black},
  identifierstyle=\color{blue},
  stringstyle=\color{red},
  numbers=left,
  numbersep=5pt,
  numberstyle=\scriptsize\color{gray},
  escapechar=@
}

\definecolor{pblue}{rgb}{0.13,0.13,1}
\definecolor{pgreen}{rgb}{0,0.5,0}
\definecolor{pred}{rgb}{0.9,0,0}
\definecolor{pgrey}{rgb}{0.46,0.45,0.48}

\lstdefinelanguage{customJava}{language=Java,
  showspaces=false,
  showtabs=false,
  breaklines=true,
  showstringspaces=false,
  breakatwhitespace=true,
  commentstyle=\color{pgreen},
  keywordstyle=\color{pblue},
  stringstyle=\color{pred},
  basicstyle=\scriptsize\ttfamily,
  numbers=left,
  numbersep=5pt,
  numberstyle=\scriptsize\color{gray},
  moredelim=[is][\textcolor{pgrey}]{\%\%}{\%\%},
  escapechar=@
}

\title{Semantic Fuzzing with \tool{}}

\author{Rohan Padhye}
\affiliation{%
  \institution{University of California, Berkeley}
  \country{USA}
}
\email{rohanpadhye@cs.berkeley.edu}

\author{Caroline Lemieux}
\affiliation{%
  \institution{University of California, Berkeley}
  \country{USA}
}
\email{clemieux@cs.berkeley.edu}

\author{Koushik Sen}
\affiliation{%
  \institution{University of California, Berkeley}
  \country{USA}
}
\email{ksen@cs.berkeley.edu}

\author{Mike Papadakis}
\affiliation{%
  \institution{University of Luxembourg}
  \country{Luxembourg}
}
\email{michail.papadakis@uni.lu}

\author{Yves Le Traon}
\affiliation{%
  \institution{University of Luxembourg}
  \country{Luxembourg}
}
\email{yves.letraon@uni.lu}

\makeatletter
\renewcommand{\ALG@beginalgorithmic}{\small}
\makeatother

\begin{abstract}

Programs expecting structured inputs often consist of both a \emph{syntactic analysis stage}, which parses raw input, and a \emph{semantic analysis stage}, which conducts checks on the parsed input and executes the core logic of the program. Generator-based testing tools in the lineage of QuickCheck are a promising way to generate random syntactically valid test inputs for these programs. 
We present \textit{\tool{}}, a technique which automatically guides QuickCheck-like random-input generators to better explore the semantic analysis stage of test programs. \tool{} converts random-input generators into deterministic \emph{parametric generators}. We present the key insight that mutations in the untyped parameter domain map to structural mutations in the input domain. \tool{} leverages program feedback in the form of code coverage and input validity to perform \emph{feedback-directed parameter search}. We evaluate \tool{} against AFL and QuickCheck on five Java programs: Maven, Ant, BCEL, Closure, and Rhino.
\tool{}
covers $1.03\times$--$2.81\times$ as many branches within the benchmarks' semantic analysis stages as baseline techniques.
Further, we find 10 new bugs in the semantic analysis stages of these benchmarks. \tool{} is the most effective technique in finding these bugs reliably and quickly, requiring at most 10 minutes on average to find each bug.
\end{abstract}



\begin{document}
\maketitle

\section{Introduction}

Programs expecting complex structured inputs often process their inputs and convert them into suitable data structures before invoking the actual functionality of the program.  For example, a build system such as Apache Maven %
first parses its input as an XML document and checks its conformance to a schema before invoking the actual build functionality.  Document processors, Web browsers, compilers and various other programs follow this same check-then-run pattern.

\begin{figure}
    \centering
    \includegraphics[width=0.9\columnwidth]{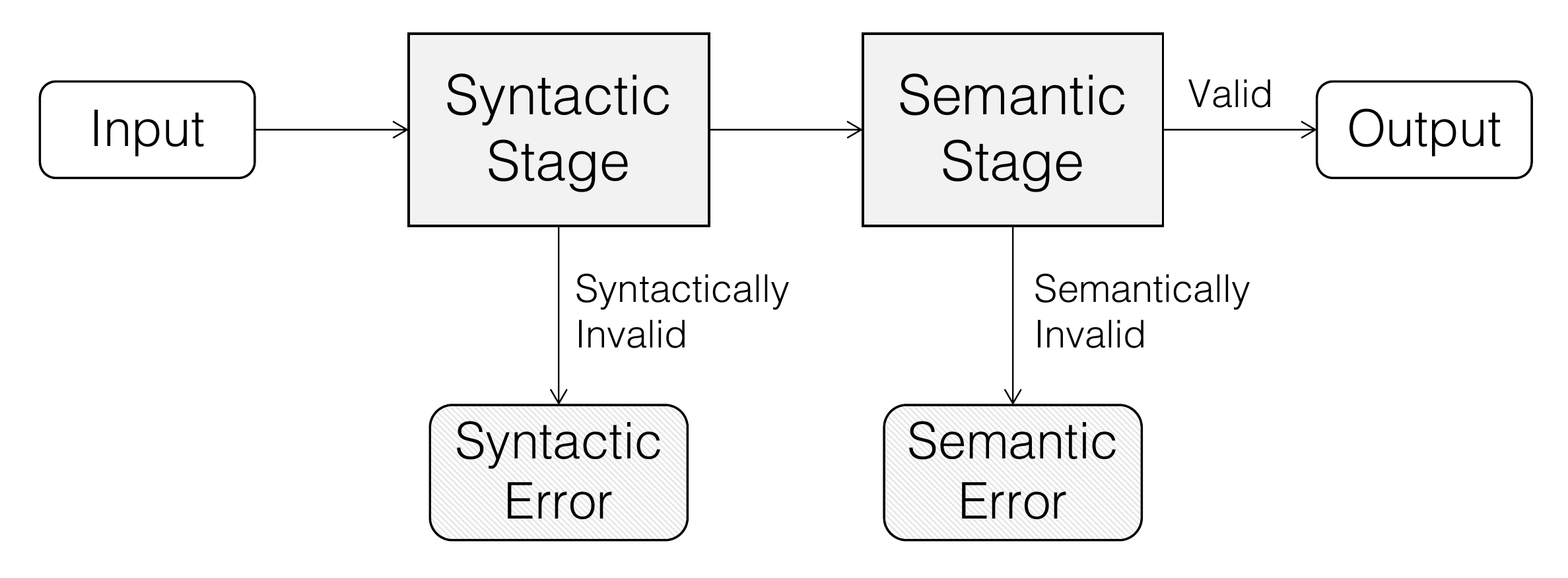}
    \caption{Inputs to a program taking structured inputs can be either syntactically or semantically invalid or just \emph{valid}.}
    \label{fig:pipeline}
\end{figure}

In general, such programs  have an input processing pipeline consisting of two stages: a syntax parser and a semantic analyzer. We illustrate this pipeline in Figure~\ref{fig:pipeline}.  The syntax parsing stage translates the raw input %
into an internal data structure that can be easily processed (e.g. an abstract syntax tree) by the rest of the program. The semantic analysis stage checks if an input satisfies certain semantic constraints (e.g. if an XML input fits a specific schema), and executes the core logic of the program. Inputs may be rejected by either stage if they are \emph{syntactically} or \emph{semantically invalid}. If an input passes both stages, we say the input is \emph{valid}.

Automatically testing such programs is challenging. The difficulty lies in synthesizing inputs that (1) satisfy complex constraints on their structure and (2) exercise a variety of code paths in the semantic analysis stages and beyond. Random input generation is a popular technique for such scenarios because it can easily scale to execute a large number of test cases. Developers can write domain-specific \emph{generators} from which random syntactically valid inputs---such as XML documents and abstract syntax trees--- can be sampled. Popularized by QuickCheck~\cite{Claessen00}, this approach has been adopted by many generator-based testing tools~\cite{Emek02, Coppit05, Arts06, Gligoric10, Papadakis11, Holler12, Amaral14, Lampropoulos12}. QuickCheck-like test frameworks are now available in many programming languages such as Java~\cite{junit-quickcheck}, PHP~\cite{eris}, Python~\cite{hypothesis}, JavaScript~\cite{jsverify}, Scala~\cite{scalacheck}, and Clojure~\cite{test.check}. Many commercial black-box fuzzing tools, such as Peach~\cite{peach},  beSTORM~\cite{bestorm}, Cyberflood~\cite{cyberflood}, and %
Codenomicon~\cite{codenomicon}, also leverage generators for network protocols or file formats. However, in order to effectively exercise the semantic analyses in the test program, the generators need to be tuned to produce inputs that are also \emph{semantically} valid. For example, the developers of CSmith~\cite{Yang11}, a tool that generates random C programs for testing compilers, spent significant effort manually tuning their generator to reliably produce valid C programs and to maximize code coverage in the compilers under test.

In this paper, we present \tool{}, a technique for \emph{automatically} guiding QuickCheck-like input generators to exercise various code paths in the semantic analysis stages of programs. 
\tool{} incorporates feedback from the test program in the form of \emph{semantic validity} of test inputs and the \emph{code coverage} achieved during test execution. The feedback is then used to generate new inputs via mutations.

\emph{Coverage-guided fuzzing} (\cgf{}) tools such as AFL~\cite{afl} and libFuzzer~\cite{libFuzzer} have gained a huge amount of popularity recently due to their effectiveness in finding critical bugs and security vulnerabilities in widely-used software systems. \cgf{} works by randomly mutating known inputs via operations such as bit flips and byte-level splicing to produce new inputs. If the mutated inputs lead to new code coverage in the test program, they are saved for subsequent mutation. Of course, such mutations usually lead to invalid syntax. Naturally, most of the bugs found by these \cgf{} tools lie in the syntax analysis stage of programs. \cgf{} tools often require many hours or days of fuzzing to discover deeper bugs~\cite{Klees18}, which makes them impractical for use in continuous integration systems with limited testing time budgets.

Our proposed technique, \tool{}, adapts the algorithm used by \cgf{} tools in order to quickly explore the \emph{semantic analysis} stages of test programs. \tool{} first converts QuickCheck-like random-input generators into deterministic \emph{parametric generators}, which map a sequence of untyped bits, called the {``parameters''}, to a syntactically valid input. The key insight in \tool{} is that \emph{bit-level} mutations on these parameters correspond to \emph{structural} mutations in the space of syntactically valid inputs. \tool{} then applies a \cgf{} algorithm on the domain of parameters, in order to guide the test-input generation towards semantic validity and increased code coverage in the semantic analysis stages.

We have integrated \tool{} into the open-source JQF framework~\cite{Padhye19-tool}: \url{https://github.com/rohanpadhye/jqf}. We evaluate \tool{} on five real-world Java benchmarks and compare it to 
AFL~\cite{afl} and (a Java port of) QuickCheck~\cite{junit-quickcheck}. %
Our results show that the \tool{} technique achieves significantly higher code coverage in the semantic analysis stage of each benchmark. Further, during our evaluation, we find 10 new bugs in the semantic analysis stages of our benchmarks. We find \tool{} to be the most effective technique for reliably and quickly triggering these \emph{semantic bugs}. For each benchmark, \tool{} discovers an input triggering every semantic bug in at most 10 minutes on average. \tool{} complements AFL, which is best suited for finding syntactic bugs.

To summarize, we make the following contributions:

\begin{itemize}
    \item We convert existing random-input generators into deterministic \emph{parametric generators}, enabling a mapping from bit-level mutations of parameters to structural mutations of inputs. %
    \item We present an algorithm that combines parametric generators with \emph{feedback-directed parameter search}, in order to effectively explore the semantic analysis stages of programs.
   
    \item We evaluate our Java-based implementation of \tool{} against AFL and QuickCheck on five real-world benchmarks to compare their effectiveness in exercising code paths and discovering new bugs within the semantic analysis stage.
\end{itemize}

\section{Background}

\subsection{Generator-based testing}
\label{sec:generators}

Generator-based testing tools~\cite{Claessen00, Emek02, Coppit05, Gligoric10, Papadakis11, Yang11, Holler12, Amaral14} allow users to write generator programs for producing inputs that belong to a specific type or format. These random-input generators are \emph{non-deterministic}, i.e., they sample a new input each time they are executed.
Figure~\ref{fig:xml-generator} shows a generator for XML documents in the \code{junit-quickcheck}~\cite{junit-quickcheck} framework, which is a Java port of QuickCheck~\cite{Claessen00}. 
When \code{generate()} is called, the generator uses the Java standard library XML DOM API to generate a random XML document. It constructs the root element of the document by invoking \code{genElement} (Line~\ref{line:gen-root}). Then, \code{genElement}
uses repeated calls to methods of \code{random} to generate the element's tag name (Line~\ref{line:gen-name}), any embedded
text (Lines~\ref{line:gen-textexist},~\ref{line:gen-textcontent}, and in \code{genString}), and the number of children  (Line~\ref{line:gen-numchildren}); it recursively calls \code{genElement} to generate each child node. We omitted code to generate attributes, but it can be done analogously.

Figure~\ref{fig:xml-ex} contains a sample test harness method \code{testProgram}, identified by the \code{@Property} annotation. This method expects a test input \code{xml} of type \code{XMLDocument}; the \code{@From} annotation indicates %
that inputs will be randomly generated using the \code{XMLGenerator.generate()} API. When invoked with a generated XML document, \code{testProgram} serializes the document (Line~\ref{line:xml-ex-read}) and invokes the \code{readModel} method (Line~\ref{line:xml-ex-parse}), which parses an input string into a domain-specific model. For example, Apache Maven parses \code{pom.xml} files into an internal Project Object Model (POM).
The model creation fails if the input XML document string does not meet certain syntactic and semantic requirements (Lines~\ref{line:synex}, \ref{line:semex}).  %
If the model creation is successful, the check at Line~\ref{line:xml-ex-assume} succeeds and the test harness invokes the method \code{runModel} at Line~\ref{line:xml-ex-assert} to test one of the core functionalities of the program under test. 

An XML generator like the one shown in Figure~\ref{fig:xml-generator} generates random syntactically valid XML inputs; therefore  Line~\ref{line:synex} in Figure~\ref{fig:xml-ex} will never be executed. However, the generated inputs may not be \emph{semantically} valid. That is, the inputs generated by the depicted XML generator do not necessarily conform to the schema expected by the application. In our example, the \code{readModel} method could throw a \code{ModelException} and cause the assumption at Line~\ref{line:xml-ex-assume} to fail. If this happens, QuickCheck simply discards the test case and tries again.  Writing generators that produce semantically valid inputs by construction is a challenging manual effort.

When we tested Apache Maven's model reader for \code{pom.xml} files using a generator similar to Figure~\ref{fig:xml-generator}, we found that only 0.09\% of the generated inputs were semantically valid. Moreover, even if the generator manages to generate semantically valid inputs, it may not generate inputs that exercise a variety of code paths in the  semantic  analysis  stage. In our experiments with Maven, the QuickCheck approach covers less than one-third of the branches in the semantic analysis stage than our proposed technique does. Fundamentally, this is because of the lack of coupling between the generators and the program under test. %

\begin{figure}
\center
\begin{minipage}{0.95\columnwidth}
\begin{lstlisting}[language=customJava,escapechar=!]
class !\textbf{XMLGenerator}! implements Generator<XMLDocument> {
    @Override // For Generator<XMLDocument>
    public XMLDocument !\textbf{generate}!(Random random) {
      XMLElement root = !\textbf{genElement}!(random, 1);  !\label{line:gen-root}!
      return new XMLDocument(root);
    }
    private XMLElement !\textbf{genElement}!(Random random, int depth) {
      // Generate element with random name
      String name = !\textbf{genString}!(random); !\label{line:gen-name}!
      XMLElement node = new XMLElement(name);
      if (depth < MAX_DEPTH) { // Ensures termination
          // Randomly generate child nodes
          int n = random.nextInt(MAX_CHILDREN); !\label{line:gen-numchildren}!
          for (int i = 0; i < n; i++) {
            node.appendChild(!\textbf{genElement}!(random, depth+1)); !\label{line:gen-child}!
          }
      }
      // Maybe insert text inside element
      if (random.nextBool()) { !\label{line:gen-textexist}!
        node.addText(!\textbf{genString}!(random)); !\label{line:gen-textcontent}!
      }
      return node;
    }
    private String !\textbf{genString}!(Random random) {
      // Randomly choose a length and characters
      int len = random.nextInt(1, MAX_STRLEN); !\label{line:gen-strlen}!
      String str = "";
      for (int i = 0; i < len; i++) {
        str += random.nextChar(); !\label{line:gen-strval}!
      }
      return str;
    }
}
\end{lstlisting}
\vspace{-10pt}
\end{minipage}
\caption{A simplified XML document generator.}
\label{fig:xml-generator}
\end{figure}

\begin{figure}
\center
\begin{minipage}{0.95\columnwidth}
\vspace{-4pt}
\begin{lstlisting}[language=customJava,escapechar=*]
@Property
void *\textbf{testProgram}*(@From(*\textbf{XMLGenerator}*.class) XMLDocument xml) {
  Model model = *\textbf{readModel}*(xml.toString()); *\label{line:xml-ex-read}*
  assume(model != null); // validity *\label{line:xml-ex-assume}*
  assert(runModel(model) == success);  *\label{line:xml-ex-assert}*
}
private Model *\textbf{readModel}*(String input) {
    try {
        return ModelReader.readModel(input);*\label{line:xml-ex-parse}*
    } catch (XMLParseException e) {
        return null; // syntax error *\label{line:synex}*
    } catch (ModelException e) {
        return null; // semantic error *\label{line:semex}*
    }
}
\end{lstlisting}
\vspace{-10pt}
\end{minipage}
\caption{A \code{junit-quickcheck} property that tests an XML-based component.}
\vspace{-0.4cm}
\label{fig:xml-ex}
\end{figure}

\subsection{Coverage-Guided Fuzzing}
\label{sec:cgf}
Algorithm~\ref{alg:afl-outline} describes coverage-guided fuzzing (\cgf{}). \cgf{} operates on raw test inputs represented as strings or byte-arrays. The algorithm maintains a set $\Q$ of important test inputs, which are used as candidates for future mutations.  $\Q$ is initialized with a user-provided set of {initial seed inputs} $I$ (Line \ref{line:afl-outline-init-queue}).  The algorithm repeatedly cycles through the elements of $\Q$ (Line \ref{line:afl-outline-traverse-queue}), each time picking an input from which to generate new inputs via mutation.  The number of new inputs to generate in this round %
(Line~\ref{line:afl-outline-energy}) is determined by
an implementation-specific heuristic.  \cgf{} generates new inputs by applying one or more random mutation operations on the base input (Line~\ref{line:afl-outline-mutate}). These mutations may include operations that combine subsets of other inputs in $\Q$. The given program is then executed with each newly generated input (Line~\ref{line:afl-outline-run-program}). The result of a test execution can either be $\textsc{Success}$ or $\textsc{Failure}$. If an input causes a test failure, it is added to the failure set $\mathcal{F}$ (Line~\ref{line:afl-outline-add-crash}).

The key to the \cgf{} algorithm is that instead of treating the test program as a black-box as QuickCheck does, %
 it instruments the test program to provide dynamic feedback in the form of code coverage for each run. The algorithm maintains in the variable \textit{totalCoverage} the set of all \emph{coverage points} (e.g. program branches) covered by the existing inputs.  If the successful execution of a generated input leads to the discovery of new coverage points (Line~\ref{line:afl-outline-interesting-criteria}), then this input is added to the set $\Q$ for subsequent fuzzing (Line~\ref{line:afl-outline-save-if-interesting}) and the newly covered coverage points are added to \textit{totalCoverage}. (Line~\ref{line:afl-outline-update-coverage}).

The whole process repeats until a time budget expires. Finally, \cgf{} returns the generated corpus of test inputs $\Q$ and failing inputs $\mathcal{F}$ (Line~\ref{line:afl-outline-return}). 
 \cgf{} can either be used as a technique to discover inputs that expose bugs---in the form of crashes or assertion violations---or to automatically generate a corpus of test inputs that cover various program features. %

\renewcommand{\algorithmicrequire}{\textbf{Input:}}
\renewcommand{\algorithmicensure}{\textbf{Output:}}

\begin{algorithm}[t]
\caption{Coverage-guided fuzzing.}\label{alg:afl-outline}
\begin{algorithmic}[1]
\algrenewcommand\algorithmicindent{1.0em}%
\Require program $p$, set of initial inputs $I$
\Ensure a set of test inputs and failing inputs
\State $\Q \gets I$ \label{line:afl-outline-init-queue}
\State $\mathcal{F} \gets \emptyset $
\State $\textit{totalCoverage} \gets \emptyset$ \label{line:afl-outline-init-cov}
\Repeat \label{line:afl-outline-begin-cycle}
\For{\textit{input} in $\Q$} \label{line:afl-outline-traverse-queue}
    \For{$1\leq i \leq \Call{numCandidates}{\textit{input}}$}\label{line:afl-outline-energy}
        \State $\textit{candidate}$ $\gets$ \Call{mutate}{\textit{input, $\Q$}}  \label{line:afl-outline-mutate}
        \State $\textit{coverage}, \textit{result} \gets \Call{run}{p, \textit{candidate}}$ \label{line:afl-outline-run-program}
        \If {$\textit{result} = \textsc{Failure}$}
            \State $\mathcal{F} \gets \mathcal{F} \cup \textit{candidate}$ \label{line:afl-outline-add-crash}
        \ElsIf {$\textit{coverage} \not\subseteq \textit{totalCoverage} $} \label{line:afl-outline-interesting-criteria}
            \State $\Q \gets \Q \cup \{\textit{candidate}\}$ \label{line:afl-outline-save-if-interesting}
            \State $\textit{totalCoverage} \gets \textit{totalCoverage} \cup \textit{coverage}$ \label{line:afl-outline-update-coverage}
        \EndIf
    \EndFor \label{line:afl-outline-mutate-end}
\EndFor
\Until{ given time budget expires} \label{line:afl-outline-repeat-until}
\State \Return $\Q, \mathcal{F}$ \label{line:afl-outline-return}
\end{algorithmic}
\end{algorithm}

A key limitation of existing \cgf{} tools is that they work without any knowledge about the syntax of the input. State-of-the-art \cgf{} tools~\cite{afl, libFuzzer, Bohme16, Rawat17, Lemieux18-FairFuzz, Chen18} treat program inputs as sequences of bytes. This choice of representation also influences the design of their mutation operations, which include bit-flips, arithmetic operations on word-sized segments, setting random bytes to random or ``interesting'' values (e.g. $0$, \code{MAX\_INT}), etc. %
These mutations are tailored towards exercising various code paths in programs that parse inputs with a compact syntax, such as parsers for media file formats, decompression routines, and network packet analyzers. \cgf{} tools have been very successful in finding  memory-corruption bugs (such as buffer overflows) in the syntax analysis stage of such programs due to incorrect handling of unexpected inputs.

Unfortunately, this approach often fails to exercise the core functions of software that expects highly structured inputs. For example, when AFL~\cite{afl} is applied on a program that processes XML input data, a typical input that it saves looks like: 
\setlength\topsep{0pt}
\begin{center}
 \texttt{<a b>ac\&\#84;a>}
 \end{center}
which exercises code paths that deal with syntax errors. In this case, an error-handling routine for unmatched start and end XML tags.
It is very difficult to generate inputs that will exercise new, interesting code paths in the semantic analysis stage of a program via these low-level mutations. Often, it is necessary to run \cgf{} tools for hours or days on end in order to find non-trivial bugs, making them impractical for use in a continuous integration setting. %

\section{Proposed Technique}
\label{sec:approach}

Our approach, \tool{}, adds the power of coverage-guided fuzzing to generator-based testing. First, \tool{} converts a random-input generator %
into an equivalent deterministic \emph{parametric generator}. 
\tool{} then uses \emph{feedback-directed parameter search} to search through the parameter space. This technique augments the \cgf{} algorithm by keeping track of code coverage achieved by valid inputs. This enables it to guide the search towards  deeper code paths in the semantic analysis stage. %

\subsection{Parametric Generators}
\label{sec:paragen}

Before defining parametric generators, let us return to the random XML generator from Figure~\ref{fig:xml-generator}.
Let us consider a particular path through this generator, concentrating on the calls to \texttt{nextInt}, \texttt{nextBool}, and \texttt{nextChar}. The following sequence of calls will be our running example (some calls ommitted for space):

\vspace{0.1cm}
\begin{adjustbox}{width=\columnwidth,center}
    \begin{tabular}{ll}
        \textbf{Call $\rightarrow $ result} & \textcolor{gray}{\textbf{Context}} \\
        \midrule
        \small$\texttt{random.nextInt(1, MAX\_STRLEN)} \rightarrow 3$ & \textcolor{gray}{\small{Root: \texttt{name} length} (Line~\ref{line:gen-strlen})} \\
        \small $\texttt{random.nextChar()} \rightarrow \texttt{`f'}$ & \textcolor{gray}{\small{Root: \texttt{name[0]} } (Line~\ref{line:gen-strval})} \\
        \small $\texttt{random.nextChar()} \rightarrow \texttt{`o'}$ & \textcolor{gray}{\small{Root: \texttt{name[1]} } (Line~\ref{line:gen-strval})} \\
        \small$\texttt{random.nextChar()} \rightarrow \texttt{`o'}$ & \textcolor{gray}{\small{Root: \texttt{name[2]} } (Line~\ref{line:gen-strval})} \\
        \small$\texttt{random.nextInt(MAX\_CHILDREN)} \rightarrow 2$ & \textcolor{gray}{\small{Root: \# children} (Line~\ref{line:gen-numchildren})} \\
        \small$\texttt{random.nextInt(1, MAX\_STRLEN)} \rightarrow 3$ & \textcolor{gray}{\small{Child 1: \texttt{name} length} (Line~\ref{line:gen-strlen})} \\
        \hfill{\vdots}\hfill & \\
        \small$\texttt{random.nextBool()} \rightarrow \text{False}$ & \textcolor{gray}{\small{Child 2: embed text? } (Line~\ref{line:gen-textexist})} \\
        \small$\texttt{random.nextBool()} \rightarrow \text{False}$ & \textcolor{gray}{\small{Root: embed text? } (Line~\ref{line:gen-textexist})} \\
    \end{tabular}
\end{adjustbox}
\vspace{0.1cm}

The XML document produced when the generator makes this sequence of calls looks like:
$$ x_1 = \texttt{<foo><bar>Hello</bar><baz /></foo>}.$$

In order to produce random typed values, the implementations of \texttt{random.nextInt}, \texttt{random.nextChar}, and \texttt{random.nextBool} rely on a pseudo-random source of \emph{untyped} bits. We call these untyped bits ``{\emph{parameters}''. The parameter sequence for the example above, annotated with the calls which consume the parameters, is:
$$\sigma_1 = \underbrace{\texttt{0000 0010}}_{\texttt{nextInt(1,.\hspace{-0.05cm}.\hspace{-0.05cm}.)}\rightarrow 3}~\underbrace{\texttt{0110 0110}}_{\texttt{nextChar()}\rightarrow\texttt{`f'}}~\dots~\underbrace{{\texttt{0000 0000}}}_{\texttt{nextBool()}\rightarrow\text{False}}.$$

For example, here the function \texttt{random.nextInt(}$a,b$\texttt{)} consumes eight bit parameters as a byte, $n$, and returns $n \,\%\, (b-a) + a$ as a typed integer. 
For simplicity of presentation, we show each \texttt{random.nextXYZ} function consuming the same number of parameters, but they can consume different numbers of parameters. %

We can now define a \textit{\textbf{parametric generator}}. A parametric generator is a function that takes a sequence of untyped parameters such as $\sigma_1$---the \emph{parameter sequence}---and produces a structured input, such as the XML $x_1$. A parametric generator can be implemented by simply replacing the underlying implementation of \code{Random} to consult not a pseudo-random source of bits but instead a fixed sequence of bits provided as the parameters.

While this is a very simple change, making generators deterministic and explicitly dependent on a fixed parameter sequence enables us to make the following two {key observations}:
\begin{enumerate}
    \item \emph{Every untyped parameter sequence corresponds to a syntactically valid input}---assuming the generator only produces syntactically valid inputs. 
    \item \emph{Bit-level mutations on untyped parameter sequences correspond to high-level structural mutations in the space of syntactically valid inputs}.
\end{enumerate}
Observation (1) is true by construction. The \texttt{random.nextXYZ} functions are implemented to produce correctly-typed values no matter what bits the pseudo-random source--or in our case, the parameters---provide. Every sequence of untyped parameter bits correspond to some execution path through the generator, and therefore every parameter sequence maps to a syntactically valid input. We describe how we handle parameter sequences that are longer or shorter than expected with the example sequences $\sigma_3$ and $\sigma_4$, respectively, below.

To illustrate observation (2), consider the following parameter sequence, $\sigma_2$, produced by mutating just a few bits of $\sigma_1$:
$$\sigma_2 = \texttt{0000 0010}~\underbrace{\texttt{01\EPS{01} 011\EPS{1}}}_{\texttt{nextChar()}\rightarrow\texttt{`W'}}~\dots~\texttt{0000 0000}.$$
As indicated by the annotation, all this parameter-sequence mutation does is change the value returned by the second call to \texttt{random.nextChar()} in our running example from \texttt{`f'} to \texttt{`W'}. So the generator produces the following test-input:
$$
x_2 = \texttt{ <\EPS{W}oo><bar>Hello</bar><baz /></\EPS{W}oo>}.
$$

Notice that this generated input is still syntactically valid, with ``\texttt{Woo}'' appearing both in the start and end tag delimiters. This is because the XML generator uses an internal DOM tree representation that is only serialized after the entire tree is generated. We get this syntactic-validity-preserving structural mutation for free, \emph{by construction}, and without modifying the underlying generators.

Mutating the parameter sequence can also result in more drastic high-level mutations. Suppose that $\sigma_1$ is mutated to influence the first call to \texttt{random.nextInt(MAX\_CHILDREN)} as follows:
$$\sigma_3 = \texttt{0000}~\dots~\underbrace{\texttt{0000  00\EPS{01}}}_{\texttt{nextInt(MAX\_CHILDREN)}\rightarrow 1}~\dots~\texttt{0000}.$$
Then the root node in the generated input will have only one child:
$$
x_3=\texttt{ <foo><bar>Hello</bar>}\textcolor{Bittersweet}{\blacksquare}\texttt{</foo>}
$$
($\textcolor{Bittersweet}{\blacksquare}$ designates the absence of \texttt{<baz />}). 
 Since the remaining values in the untyped parameter sequence are the same, the first child node in $x_3$---\code{<bar>Hello</bar>}---is identical to the one in $x_1$. The parametric generator thus enables a structured mutation in the DOM tree, such as deleting a sub-tree, by simply changing a few values in the parameter sequence. Note that this change results in fewer \texttt{random.nextXYZ} calls by the generator; the unused parameters in the tail of the parameter sequence will simply be ignored by the parametric generator.

For our final example, suppose $\sigma_1$ is mutated as follows:
$$\sigma_4 = \texttt{0000 0011}~\dots~\underbrace{\texttt{0000 000\EPS{1}}}_{\texttt{nextBool()}\rightarrow\text{True}} \underbrace{\texttt{0000 0000}}_{\texttt{nextInt(1,.\hspace{-0.05cm}.\hspace{-0.05cm}.)}\rightarrow 1}
.$$
Notice that after this mutation, the last 8 parameters are consumed by \texttt{nextInt} instead of by \texttt{nextBool} (ref. $\sigma_1$). But, note that \texttt{nextInt} still returns a valid typed value even though the parameters were originally consumed by \texttt{nextBool}.

At the input level, this modifies the call sequence so that the decision to embed text in the second child of the document becomes True. Then, the last parameters are used by \texttt{nextInt} to choose an embedded text length of 1 character.
However, one problem remains: to generate the content of the embedded text, the generator needs more parameter values than $\sigma_4$ contains. In \tool{}, we deal with this by appending pseudo-random values to the end of the parameter sequence on demand. We use a fixed random seed to maintain determinism. For example, suppose the sequence is extended as:
$$\sigma_4' = \texttt{0000}~\dots~\texttt{ 000\EPS{1} 0000 0000}
~\underbrace{\EPS{\texttt{0100 1100}}}_{\texttt{nextChar()}\rightarrow \texttt{`H'}}~\underbrace{\EPS{\texttt{0000 0000}}}_{\texttt{nextBool()}\rightarrow \text{False}}
$$
Then the parametric generator would produce the test-input:
$$
x_4=\texttt{ <foo><bar>Hello</bar><baz>\EPS{H}</baz></foo>}.
$$

\subsection{Feedback-directed Parameter Search}

\label{sec:algorithm}

\begin{algorithm}[t]
\caption{The \tool{} algorithm, pairing parametric generators with feedback-directed parameter search. Additions to Algorithm~\ref{alg:afl-outline} highlighted in grey.}\label{alg:zest}
\begin{algorithmic}[1]
\algrenewcommand\algorithmicindent{1.0em}%
\Require program $p$,  generator $q$  
\Ensure a set of test inputs and failing inputs
\State $\Q \gets \colorbox{gray!30}{\{\Call{random}{} \}}$ \label{line:jqf-outline-init-queue} 
\State $\mathcal{F} \gets \emptyset $
\State $\textit{totalCoverage} \gets \emptyset$\label{line:jqf-outline-init-cov} \newline
\hspace*{-\fboxsep}\colorbox{gray!30}{\parbox{\linewidth}{
\State $\textit{validCoverage} \gets \emptyset$ \label{line:jqf-outline-init-valid-cov}
\State $g \gets \Call{makeParametric}{q}$  \label{line:jqf-outline-pargen} }}
\Repeat \label{line:jqf-outline-begin-cycle}
\For{\textit{param} in $\Q$} \label{line:jqf-outline-traverse-queue}
    \For{$1\leq i \leq \Call{numCandidates}{\textit{param}}$}\label{line:jqf-outline-energy}
        \State $\textit{candidate}$ $\gets$ \Call{mutate}{\textit{param, $\Q$}}  \label{line:jqf-outline-mutate}
        \newline
     \hspace*{-\fboxsep}\colorbox{gray!30}{\parbox{\linewidth}{
        \State $\textit{input} \gets g(\textit{candidate})$ \label{line:jqf-outline-paragen}
        }}
        \State $\textit{coverage}, \textit{result} \gets \Call{run}{p, \colorbox{gray!30}{\textit{input}}}$ \label{line:jqf-outline-run-program}
        \If {$\textit{result} = \textsc{Failure}$}
            \State $\mathcal{F} \gets \mathcal{F} \cup \textit{candidate}$
        \Else 
            \If{$\textit{coverage} \not\subseteq \textit{totalCoverage} $} \label{line:jqf-outline-interesting-criteria}
                \State $\Q \gets \Q \cup \{\textit{candidate}\}$ \label{line:jqf-outline-save-if-interesting}
                \State $\textit{totalCoverage} \gets \textit{totalCoverage} \cup \textit{coverage}$ \label{line:jqf-outline-update-coverage}
            \EndIf
            \newline
            \hspace*{-\fboxsep}\colorbox{gray!30}{\parbox{\linewidth}{
            \If {$\textit{result} = \textsc{Valid}$ and $\textit{coverage} \not\subseteq \textit{validCoverage} $} \label{line:jqf-outline-interesting-valid}
                \State $\Q \gets \Q \cup \{\textit{candidate}\}$
                \State $\textit{validCoverage} \gets \textit{validCoverage} \cup \textit{coverage}$ \label{line:jqf-outline-update-valid}
            \EndIf
            }}
       \EndIf

    \EndFor \label{line:jqf-outline-mutate-end}
\EndFor
\Until{ given time budget expires} \label{line:jqf-outline-repeat-until}
\State \Return \colorbox{gray!30}{$g(\Q),~g(\mathcal{F})$} \label{line:jqf-outline-return}
\end{algorithmic}
\end{algorithm}

Algorithm~\ref{alg:zest} shows the \tool{} algorithm, which guides parametric generators to produce inputs that get deeper into the semantic analysis stage of programs using \emph{feedback-directed parameter search}. The Zest algorithm resembles Algorithm~\ref{alg:afl-outline}, but acts on parameter sequences rather than the raw inputs to the program. It also extends the \cgf{} algorithm by keeping track of the coverage achieved by \emph{semantically valid inputs}.
We highlight the differences between Algorithms~\ref{alg:zest} and~\ref{alg:afl-outline} in grey.

Like Algorithm~\ref{alg:afl-outline}, \tool{} is provided a program under test $p$. Unlike Algorithm~\ref{alg:afl-outline} which assumes seed inputs, the set of parameter sequences is initialized with a random sequence (Line~\ref{line:jqf-outline-init-queue}).
Additionally, \tool{} is provided a generator $q$, which it automatically converts to a parametric generator $g$ (Line~\ref{line:jqf-outline-pargen}). In an abuse of notation, we use $g(S)$ to designate the set of inputs generated by running $g$ over the parameter sequences in $S$, i.e. $g(S)= \{g(s) : s\in S\}$. 

Along with $\textit{totalCoverage}$, which maintains the set of coverage points in $p$ covered by \emph{all} inputs in $g(\Q)$, \tool{}  also maintains $\textit{validCoverage}$, the set of coverage points covered by the (semantically) valid inputs in $g(\Q)$. This is  initialized at Line~\ref{line:jqf-outline-init-valid-cov}.

New parameter sequences are generated using standard \cgf{} mutations at Line~\ref{line:jqf-outline-mutate}; see Section~\ref{sec:implementation} for details. New inputs are generated by running the sequences through the parametric generator (Line~\ref{line:jqf-outline-paragen}). The program $p$ is then executed on each input. During the execution, in addition to code-coverage and failure feedback, the algorithm records in the variable $\textit{result}$ whether the input is valid or not. In particular,  $\textit{result}$  can be any of $\{\textsc{Valid}, \textsc{Invalid}, \textsc{Failure} \}$. An input is considered invalid if it leads to a violation of any assumption in the test harness (e.g. Figure~\ref{fig:xml-ex} at Line~\ref{line:xml-ex-assume}), which is how we capture application-specific semantic validity.%

As in Algorithm~\ref{alg:afl-outline}, a newly generated parameter sequence is added to the set $\Q$ at Lines~\ref{line:jqf-outline-interesting-criteria}--\ref{line:jqf-outline-update-coverage} of  Algorithm~\ref{alg:zest} if the corresponding input produces new code coverage. Further, if the corresponding input is \emph{valid} and  covers a coverage point that has not been exercised by \emph{any previous valid input}, then the parameter sequence is added $\Q$ and the cumulative valid coverage variable \textit{validCoverage} is updated at Lines~\ref{line:jqf-outline-interesting-valid}--\ref{line:jqf-outline-update-valid}. Adding the parameter sequence to $\Q$ under this new condition ensures that \tool{} keeps mutating valid inputs that exercise  core program functionality. %
We hypothesize that this biases the search towards generating even more valid inputs and in turn increases code coverage in the semantic analysis stage. %

As in Algorithm~\ref{alg:afl-outline}, the testing loop repeats until a time budget expires. Finally, \tool{} returns the corpus of generated test inputs $g(\Q)$ and failing inputs $g(\mathcal{F})$.%

\section{Implementation}
\label{sec:implementation}

\tool{} is implemented on top of the open-source JQF platform~\cite{Padhye19-tool}, which provides a framework for specifying algorithms for feedback-directed fuzz testing of Java programs. JQF dynamically instruments Java classes in the program under test using the ASM bytecode-manipulation framework~\cite{asm} via a \code{javaagent}. The instrumentation allows JQF to observe  code coverage events, e.g. the execution of program branches and invocation of virtual method calls.

Fuzzing ``guidances'' can plug into JQF to provide inputs and register callbacks for listening to code coverage events. JQF originally shipped with \code{AFLGuidance} and \code{NoGuidance}, which we use in our evaluation in Section~\ref{sec:eval}. \code{AFLGuidance} uses a proxy program to exchange program inputs and coverage feedback with the external AFL tool; the overhead of this inter-process communication is a negligible fraction of the test execution time. \code{NoGuidance} randomly samples inputs from \code{junit-quickcheck}~\cite{junit-quickcheck} generators without using coverage feedback. 
We implement \code{ZestGuidance} in JQF, which biases these generators using Algorithm~\ref{alg:zest}. 

\code{junit-quickcheck} provides a high-level API for making random choices in the generators, such as generating random integers, time durations, and selecting random items from a collection.
All of these methods indirectly rely on the underlying JDK method \code{java.util.Random.next(int bits)}, which returns bits from a pseudo-random stream. \tool{} replaces this pseudo-random stream with stored parameter sequences, which are extended on-demand.

Since \code{java.util.Random} polls byte-sized chunks from its underlying stream of pseudo-random bits, \tool{} performs mutations on the parameter sequences (Algorithm~\ref{alg:zest}, Line~\ref{line:jqf-outline-mutate}) at the \emph{byte-level}. The basic mutation procedure is as follows: (1) choose a random number $m$ of mutations to perform sequentially on the original sequence, (2) for each mutation, choose a random length $\ell$ of bytes to mutate and an offset $k$ at which to perform the mutation, and (3) replace the bytes from positions $[k, k + \ell)$ with $\ell$ randomly chosen bytes. The random numbers $m$ and $\ell$ are chosen from a geometric distribution, which mostly provides small values without imposing an upper bound. We set the mean of this distribution to $4$, since 4-byte \code{int}s are the most commonly requested random value. %

\section{Evaluation}
\label{sec:eval}

We evaluate \tool{} by measuring its effectiveness in testing the semantic analysis stages of five benchmark programs. We compare \tool{} with two baseline techniques: AFL and \code{junit-quickcheck} (referred to as simply QuickCheck hereon). AFL is known to excel in exercising the syntax analysis stage via coverage-guided fuzzing of raw input strings. We use version 2.52b, with ``FidgetyAFL'' configuration, which was found to match the performance of AFLFast~\cite{FidgetyAFL}. QuickCheck uses the same generators as \tool{} but only performs random sampling without any feedback from the programs under test. Specifically, we evaluate the three techniques on two fronts: (1) the amount of code coverage achieved in the semantic analysis stage after a fixed amount of time, and (2) their effectiveness in triggering bugs in the semantic analysis stage. 

\paragraph{Benchmarks}
We use the following five real-world Java benchmarks as test programs for our evaluation: 
\begin{enumerate}
    \item Apache Maven~\cite{maven} (99k LoC): The test reads a \code{pom.xml} file and converts it into an internal \code{Model} structure. The test driver is similar to the one in Figure~\ref{fig:xml-ex}. An input is valid if it is a valid XML document conforming to the POM schema.
    \item Apache Ant~\cite{ant} (223k LoC): Similar to Maven, this test reads a \code{build.xml} file and populates a \code{Project} object. An input is considered valid if it is a valid XML document and if it conforms to the schema expected by Ant.
    \item Google Closure~\cite{closure} (247k LoC) statically optimizes JavaScript code. The test driver invokes the \code{Compiler.compile()} on the input with the \code{SIMPLE\_OPTIMIZATIONS} flag, which enables constant folding, function inlining, dead-code removal, etc.. An input is valid if Closure returns without error.
    \item Mozilla Rhino~\cite{rhino} (89k LoC) compiles JavaScript to Java bytecode. The test driver invokes \code{Context.compileString()}. An input is valid if Rhino returns a compiled script.
    \item Apache's Bytecode Engineering Library (BCEL)~\cite{bcel} (61k LoC) provides an API to parse, verify and manipulate Java bytecode. Our test driver takes as input a \code{.class} file and uses the \code{Verifier} API to perform 3-pass verification of the class file according to the Java 8 specification~\cite{jvm8-spec}. An input is valid if BCEL finds no errors up to Pass 3A verification. %
\end{enumerate}

\newcommand\pkgname[1]{\code{#1}}

\begin{table*}[t]
    \centering
        \caption{Description of benchmarks with prefixes of class/package names corresponding to syntactic and semantic analyses.}
    \begin{tabular}{lllll}\toprule
        \textbf{Name} & \textbf{Version} 
        & \textbf{Syntax Analysis Classes} & \textbf{Semantic Analysis Classes} \\ \midrule
         Maven & 3.5.2
         & \pkgname{org/codehaus/plexus/util/xml} & \pkgname{org/apache/maven/model}  \\
         Ant & 1.10.2
         & \pkgname{com/sun/org/apache/xerces} & \pkgname{org/apache/tools/ant}\\
         Closure & v20180204
         & \pkgname{com/google/javascript/jscomp/parsing} & \pkgname{com/google/javascript/jscomp/[A-Z]}\\
         Rhino & 1.7.8
         & \pkgname{org/mozilla/javascript/Parser} & \pkgname{org/mozilla/javascript/(optimizer|CodeGenerator)} \\
         BCEL & 6.2
         & \pkgname{org/apache/bcel/classfile} & \pkgname{org/apache/bcel/verifier}\\ \bottomrule
    \end{tabular}
    \label{table:benchmarks}
    \vspace{-0.0cm}
\end{table*}

\paragraph{Experimental Setup} We make the following design decisions:

\begin{itemize}
    \item \textbf{Duration}: We run each test-generation experiment for \emph{3 hours}. Researchers have used various timeouts to evaluate random test generation tools, from 2 minutes~\cite{Pacheco07, Fraser14} to 24 hours~\cite{Bohme16, Klees18}. %
    We chose 3 hours as a middle ground. %
    Our experiments justify this choice, as we found that  semantic coverage plateaued after 2 hours in almost all experiments. Specifically, the number of semantic branches covered by \tool{} increased by less than 1\% in the last hour of the runs.
    \item \textbf{Repetitions}: Due to the non-deterministic nature of random testing, the results may vary across multiple repetitions of each experiment. We therefore run each experiment 20 times and report statistics across the 20 repetitions.
    \item \textbf{Seeds and Dictionaries}: 
    To bootstrap AFL, we need to provide some initial seed inputs. There is no single best strategy for selecting initial seeds~\cite{Rebert14}. 
    Researchers have found success using varying strategies ranging from large seed corpora to single empty files~\cite{Klees18}. In our evaluation, we provide AFL one valid seed input per benchmark that covers various domain-specific semantic features.  For example, in the Closure and Rhino benchmarks, we use the entire React.JS library~\cite{react} as a seed.
    We also provide AFL with \emph{dictionaries} of benchmark-specific strings (e.g. keywords, tag names) to inject into inputs during mutation. The generator-based tools \tool{} and QuickCheck do not rely on meaningful seeds.

    \item \textbf{Generators}: \tool{} and QuickCheck use hand-written input generators. For Maven and Ant, we use an XML document generator similar to Figure~\ref{fig:xml-generator}, of around 150 lines of Java code. It generates strings for tags and attributes by randomly choosing strings from a list of string literals scraped from class files in Maven and Ant. For Closure and Rhino, we use a generator for a subset of JavaScript that contains about 300 lines of Java code. The generator produces strings that are syntactically valid JavaScript programs. %
    Finally, the BCEL generator (\textasciitilde{}500 LoC) uses the BCEL API to generate \code{JavaClass} objects with randomly generated fields, attributes and methods with randomly generated bytecode instructions. All generators were written by one of the authors of this paper in less than two hours each. Although these generators produce syntactically valid inputs, no effort was made to produce semantically valid inputs; doing so can be a complex and tedious task~\cite{Yang11}.
\end{itemize}

The generators, seeds, and dictionaries have been made publicly available at \url{https://goo.gl/GfLRzA}. All experiments are conducted on a machine with Intel(R) Core(TM) i7-5930K 3.50GHz CPU and 16GB of RAM running Ubuntu 18.04.

\paragraph{Syntax and Semantic Analysis Stages in Benchmarks} \tool{} is specifically designed to exercise the semantic analysis stages of programs. To evaluate \tool{}'s effectiveness in this regard, we manually identify the components of our benchmark programs which correspond to syntax and semantic analysis stages. Table~\ref{table:benchmarks} lists prefix patterns that we match on the fully-qualified names of classes in our benchmarks to classify them in either stage. Section~\ref{sec:eval-coverage} evaluates the code coverage achieved within the classes identified as belonging to the semantic analysis stage. Section~\ref{sec:eval-bugs} evaluates the bug-finding capabilities of each technique for bugs that arise in the semantic analysis classes. Section~\ref{sec:discussion} discusses some findings in the \emph{syntax} analysis classes, whose testing is outside the scope of \tool{}.

\subsection{Coverage of Semantic Analysis Classes}
\label{sec:eval-coverage}

\begin{figure}[t]
    \centering
    \begin{subfigure}[t]{0.235\textwidth}
        \includegraphics[width=\columnwidth]{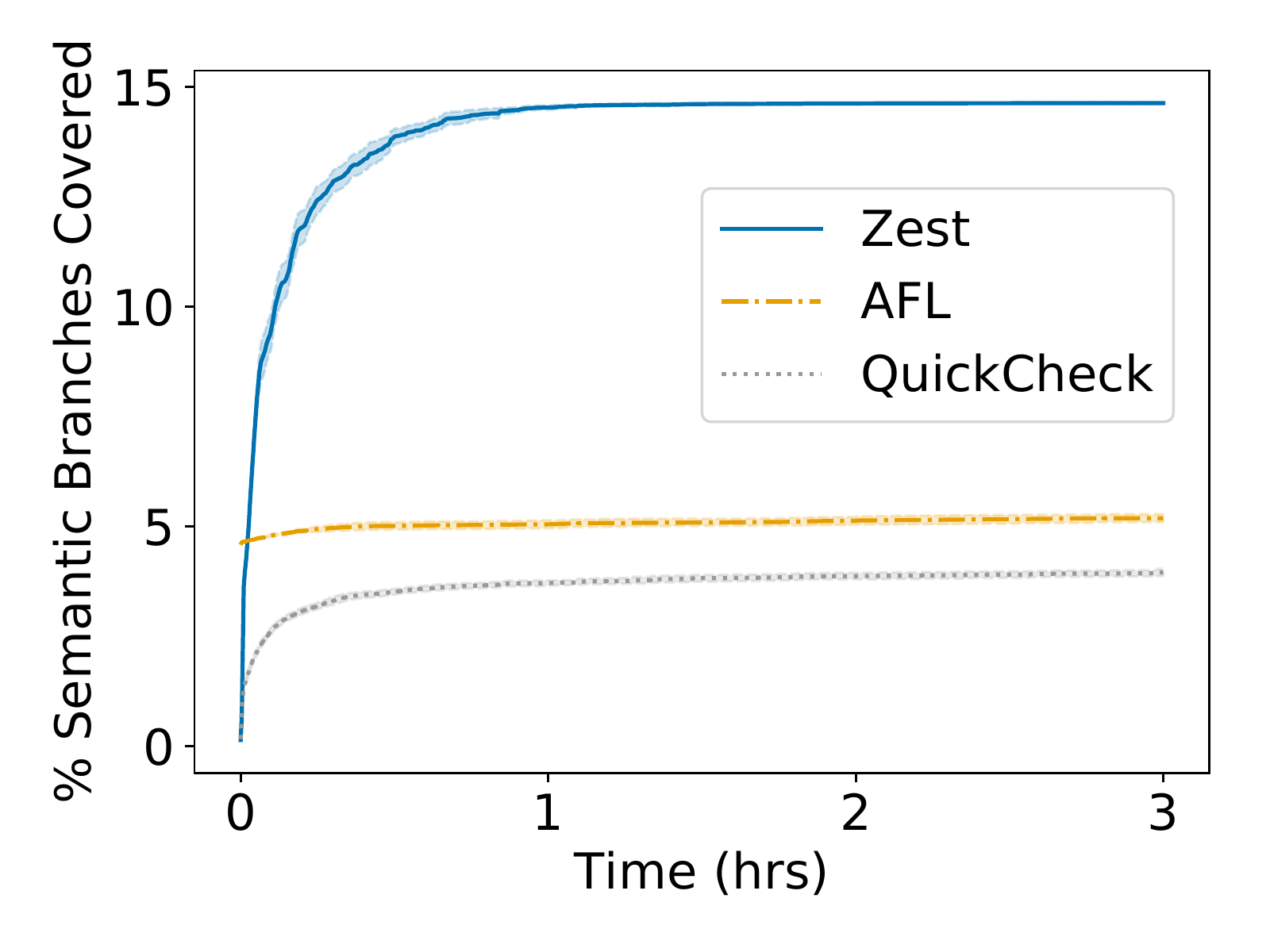}
        \caption{maven}
        \label{fig:coverage_graphs_maven}
    \end{subfigure}
    \begin{subfigure}[t]{0.235\textwidth}
        \includegraphics[width=\columnwidth]{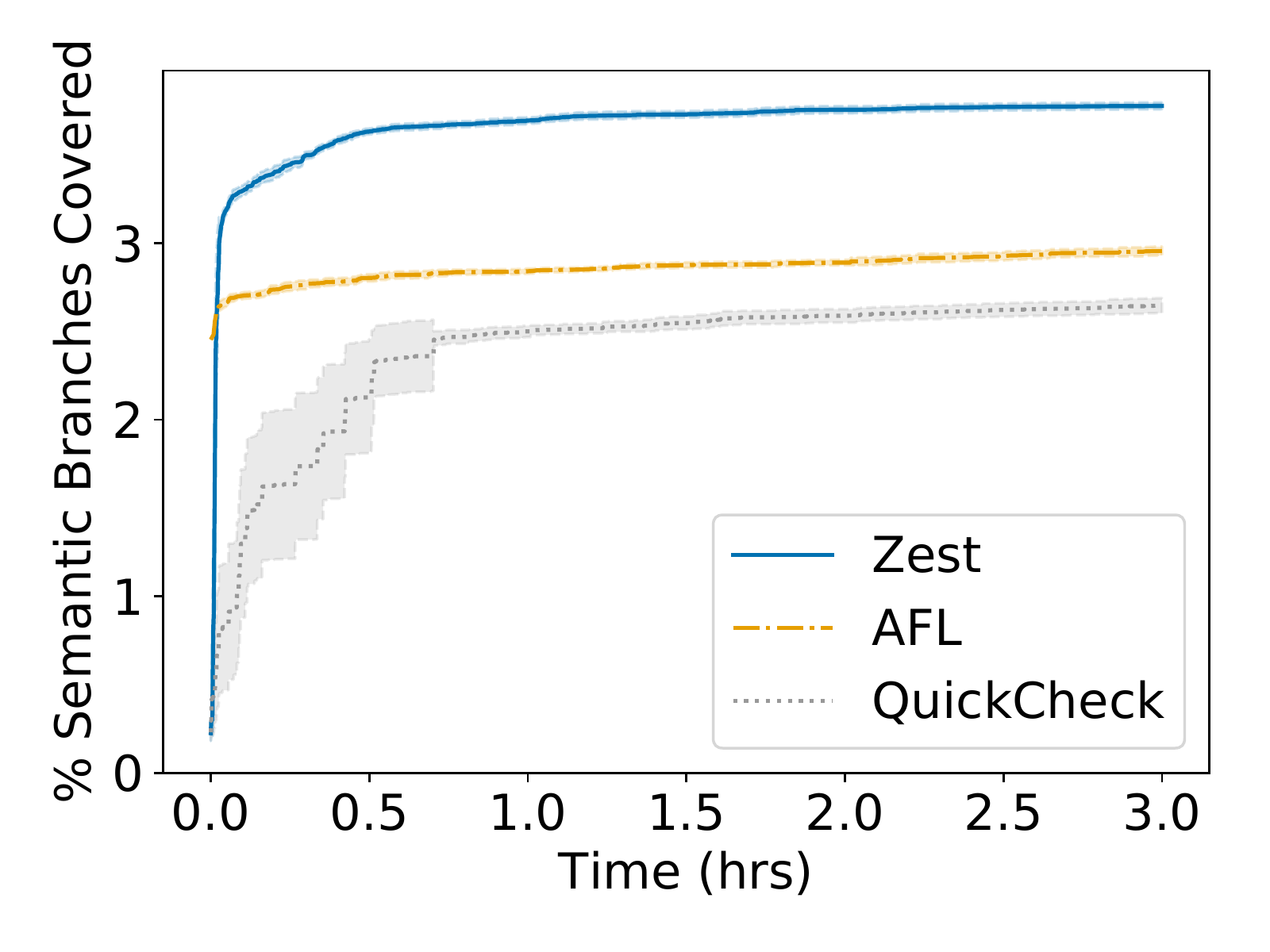}
        \caption{ant}
        \label{fig:coverage_graphs_ant}
    \end{subfigure}
    \begin{subfigure}[t]{0.235\textwidth}
        \includegraphics[width=\columnwidth]{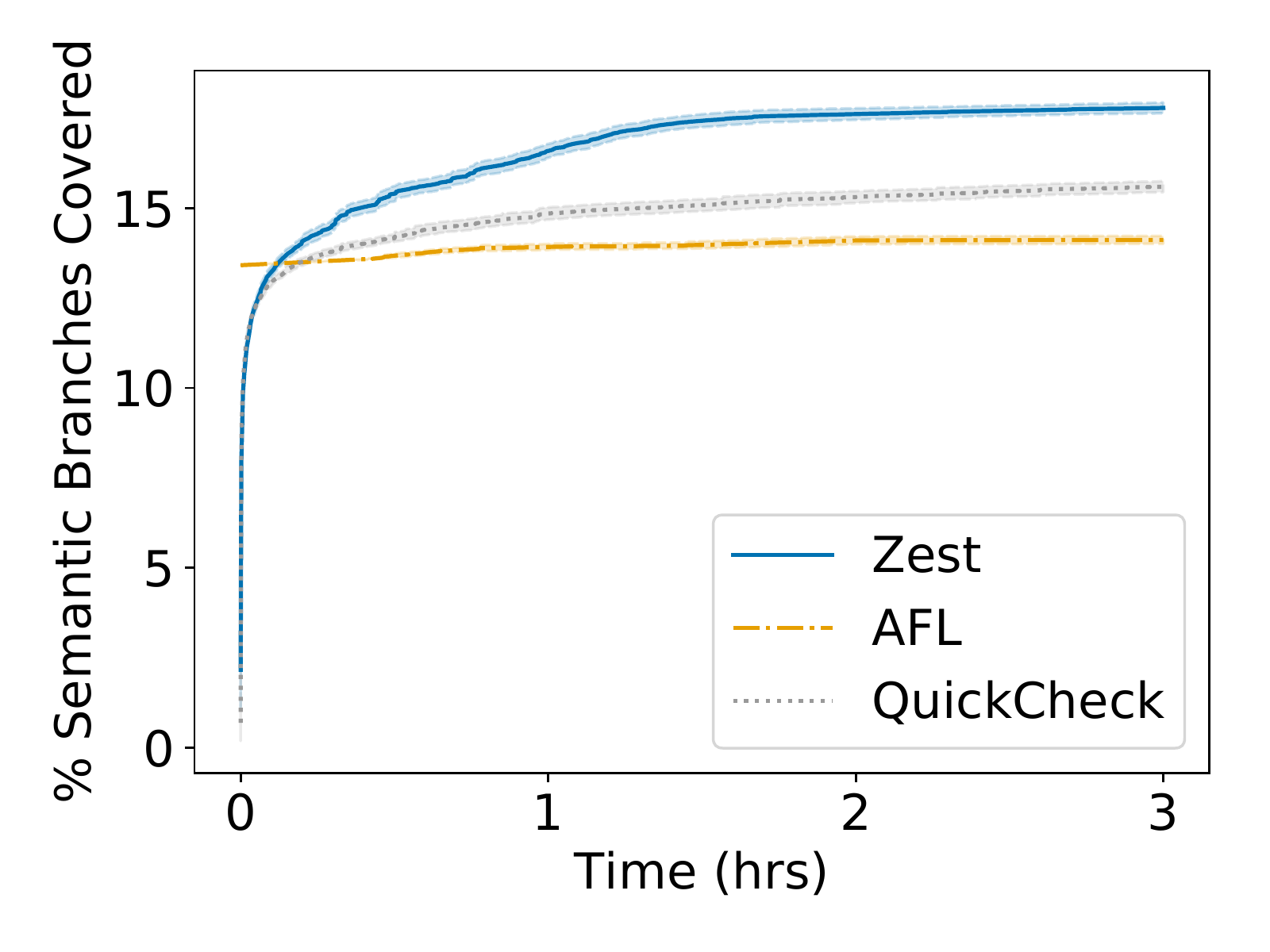}
        \caption{closure}
        \label{fig:coverage_graphs_closure}
    \end{subfigure}
    \begin{subfigure}[t]{0.235\textwidth}
        \includegraphics[width=\columnwidth]{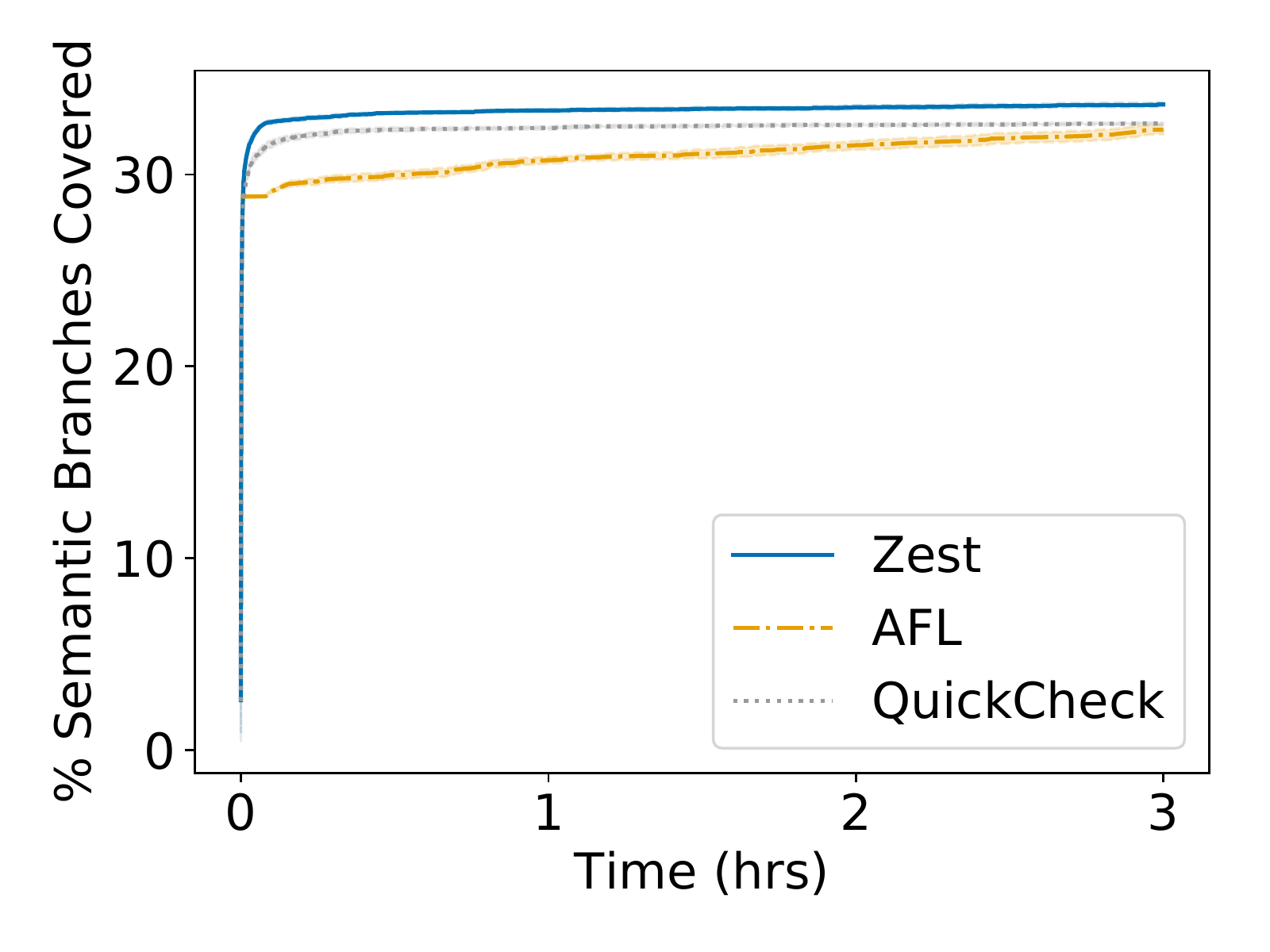}
        \caption{rhino}
        \label{fig:coverage_graphs_rhino}
    \end{subfigure}
    \begin{subfigure}[t]{0.235\textwidth}
        \includegraphics[width=\columnwidth]{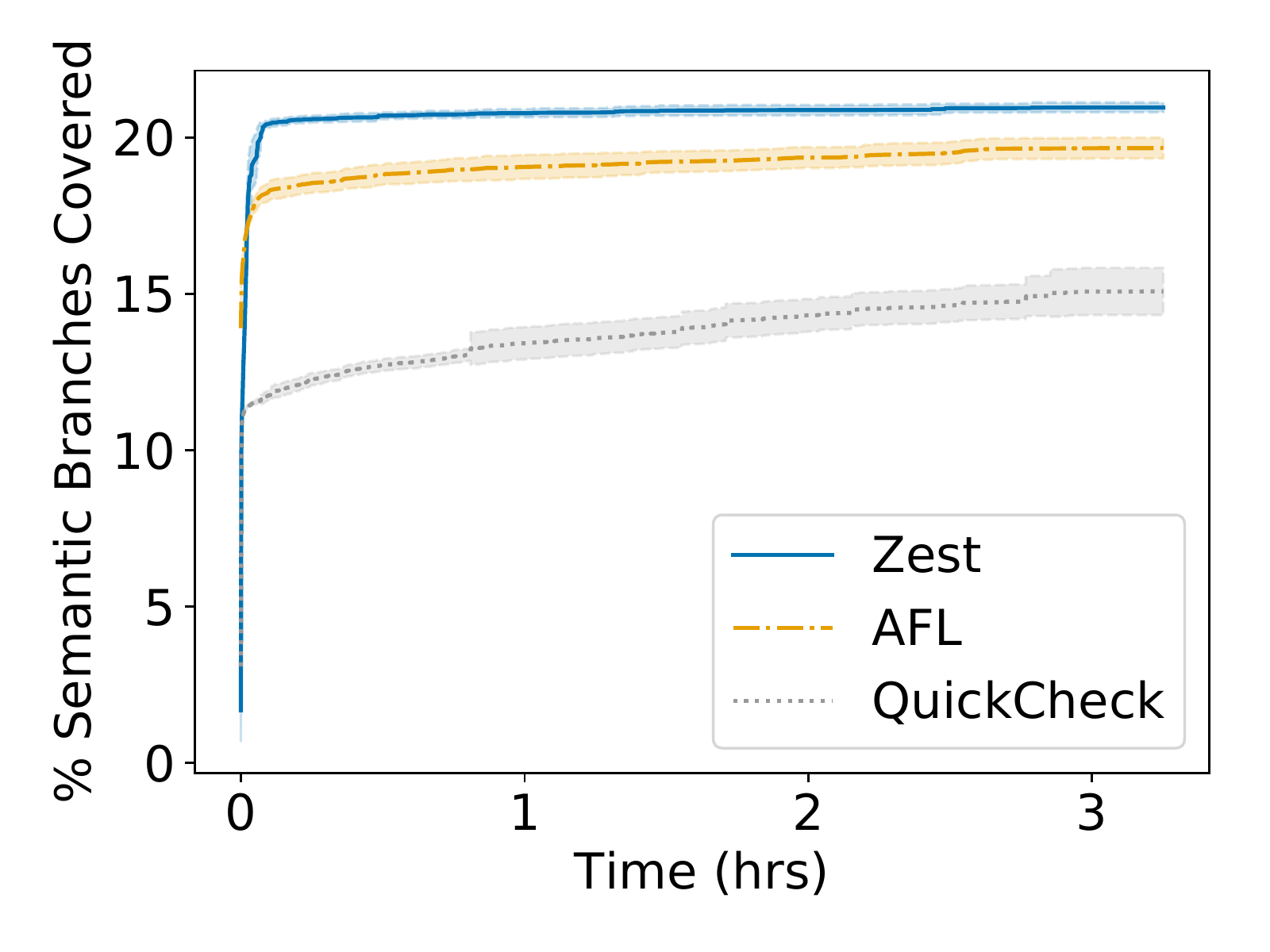}
        \caption{bcel}
        \label{fig:coverage_graphs_bcel}
    \end{subfigure}
    \caption{Percent coverage of \emph{all} branches in semantic analysis stage of the benchmark programs. Lines designate means and shaded regions 95\% confidence intervals.}
    \label{fig:coverage_graphs}
\end{figure}

Instead of relying on our own instrumentation, we use a third party tool, the widely used Eclemma-JaCoCo~\cite{jacoco} library, for measuring code coverage in our Java benchmarks. Specifically, we measure \emph{branch coverage} within the semantic analysis classes from Table~\ref{table:benchmarks}; we refer to these branches as \emph{semantic branches} for short.

To approximate the coverage of the semantic branches covered via the selected test drivers, we report the percentage of total semantic branches covered. Note, however, that this is a \emph{conservative}, i.e. low, estimate. This is because the total number of semantic branches includes some branches not reachable from the test driver. 
We make this approximation as it is not feasible to statically determine the number of branches reachable from a given entry point, especially in the presence of virtual method dispatch. We expect the percent of semantic branches reachable from our test drivers to be much lower than 100\%; therefore, the relative differences between coverage are more important than the absolute percentages.

Figure~\ref{fig:coverage_graphs} plots the semantic branch coverage achieved by each of \tool{}, AFL, and QuickCheck on the five benchmark programs across the 3-hour-long runs. In the plots, solid lines designate means and shaded areas designate 95\% confidence intervals across the 20 repetitions. Interestingly, the variance in coverage is quite low for all techniques except QuickCheck.  Since AFL is initialized with valid seed inputs, its initial coverage is non-zero; nonetheless, it is quickly overtaken by \tool{}, usually within the first 5 minutes.

\tool{} significantly outperforms baseline techniques in exercising branches within the semantic analysis stage, achieving statistically significant increases for all benchmarks. \tool{} covers as much as $2.81\times$ as many semantic branches covered by the best baseline technique for Maven (Figure~\ref{fig:coverage_graphs_maven}).  
When looking at our Javascript benchmarks, we see that \tool{}'s advantage over QuickCheck is more slight in Rhino (Figure~\ref{fig:coverage_graphs_ant}) than in Closure (Figure~\ref{fig:coverage_graphs_closure}).
This may be because Closure, which performs a variety of static code optimizations on JavaScript programs, has many input-dependent paths. Rhino, on the other hand, directly compiles valid JavaScript to JVM bytecode, and thus has fewer input-dependent paths for \tool{} to discover through feedback-driven parameter search.

Note that in some benchmarks AFL has an edge in coverage over QuickCheck (Figure~\ref{fig:coverage_graphs_maven},~\ref{fig:coverage_graphs_ant},~\ref{fig:coverage_graphs_bcel}), and vice-versa (Figure~\ref{fig:coverage_graphs_closure},~\ref{fig:coverage_graphs_rhino}). For BCEL, this may be because the input format is a compact syntax, on which AFL generally excels. The difference between the XML and JavaScript benchmarks may be related to the ability of randomly-sampled inputs from the generator to achieve broad coverage. It is much more likely for a random syntactically valid JavaScript program to be semantically valid than a random syntactically valid XML document to be a valid POM file, for example. The fact that \tool{} dominates the baseline approaches in all these cases suggests that it is more robust to generator quality than QuickCheck.

\newcommand*\synctacticBug[1]{\tikz[baseline=(char.base)]{
            \node[shape=circle,draw,inner sep=1pt] (char) {\scriptsize #1};}}
            
\newcommand*\semanticBug[1]{\tikz[baseline=(char.base)]{
            \node[shape=circle,draw,inner sep=1pt,fill=black!0] (char) {\scriptsize #1};}}

\newcommand\bugA{\synctacticBug{A}} %
\newcommand\bugS{\synctacticBug{S}} %
\newcommand\bugT{\synctacticBug{T}} %

\newcommand\bugB{\semanticBug{B}} %

\newcommand\bugC{\semanticBug{C}} %
\newcommand\bugD{\semanticBug{D}} %
\newcommand\bugU{\semanticBug{U}} %

\newcommand\bugE{\synctacticBug{E}} %
\newcommand\bugF{\semanticBug{F}} %
\newcommand\bugG{\semanticBug{G}} %
\newcommand\bugH{\semanticBug{H}} %
\newcommand\bugI{\synctacticBug{I}} %
\newcommand\bugJ{\semanticBug{J}} %

\newcommand\bugK{\semanticBug{K}}   %
\newcommand\bugL{\synctacticBug{L}}   %
\newcommand\bugM{\semanticBug{M}} %
\newcommand\bugN{\semanticBug{N}} %
\newcommand\bugO{\semanticBug{O}}  %
\newcommand\bugP{\semanticBug{P}}  %
\newcommand\bugQ{\semanticBug{Q}}  %
\newcommand\bugR{\semanticBug{R}}   %

\tikzset{ brokenrect/.style={
    append after command={
      \pgfextra{
      \path[draw,#1]
       decorate[decoration={zigzag,segment length=0.3em, amplitude=.3mm}]
       {(\tikzlastnode.north east)--(\tikzlastnode.south east)}      
        -- (\tikzlastnode.south west)|-cycle;
        }}}}

\newcommand\mtf[2]{\begin{tikzpicture}[xscale=0.05, yscale=0.20]
\filldraw [fill=black!10, draw=black] (0,0) rectangle ++(#2/100, 1);
\end{tikzpicture} (#2 sec) & #1\%} 
\newcommand\ntf{\begin{tikzpicture}[xscale=0.05, yscale=0.20]
\node [brokenrect={draw=black, fill=black!10}, inner xsep=100] {};
 \end{tikzpicture} \,\ding{55} & 0\%} 

\newcommand{\rsquare}[1]{%
  \tikz[baseline=(char.base)]\node[anchor=south west, draw,rectangle, rounded corners, inner sep=2pt](char){#1} ;}
\newcommand\winner[1]{\rsquare{\textbf{#1}}}

\begin{table*}[t]
\caption{The 10 new bugs found in the semantic analysis stages of benchmark programs. The tools \tool{}, AFL, and QuickCheck (QC) are evaluated on the \emph{mean time to find} (MTF) each bug across the 20 repeated experiments of 3 hours each as well as the \emph{reliability} of this discovery, which is the percentage of the 20 repetitions in which the bug was triggered at least once. For each bug, the highlighted tool is statistically significantly more effective at finding the bug than unhighlighted tools.}
\begin{adjustbox}{width=\textwidth,center}
\footnotesize
\begin{tabular}{llllc}
\toprule
     \textbf{Bug ID} & \textbf{Exception} & \textbf{Tool} & \textbf{Mean Time to Find (shorter is better)} &
     \textbf{Reliability} \\%& \textbf{Status}\\
     \midrule
     \multirow{3}{*}{ant \bugB} & \multirow{3}{*}{\code{IllegalStateException}} &  
     \winner{\tool{}} & \mtf{100}{99.45} \\%& \multirow{3}{*}{Fixed} \\ %
     & & AFL & \mtf{10}{6369.5} \\
     & & QC & \mtf{10}{1208.0}  \\
     \midrule
     \multirow{3}{*}{closure \bugC} & \multirow{3}{*}{\code{NullPointerException}} &  
     \winner{\tool{}} & \mtf{100}{8.8} \\%& \multirow{3}{*}{Fixed} \\ %
     & & AFL & \mtf{20}{5496.25} \\
     & & \winner{QC} & \mtf{100}{8.8}  \\
     \midrule
     \multirow{3}{*}{closure \bugD} & \multirow{3}{*}{\code{RuntimeException}} &  
     \winner{\tool{}} & \mtf{60}{460.42} \\%& \multirow{3}{*}{Confirmed} \\ %
     & & AFL & \ntf \\
     & & QC & \ntf  \\
     \midrule
     \multirow{3}{*}{closure \bugU} & \multirow{3}{*}{\code{IllegalStateException}} &  
     \winner{\tool{}} & \mtf{5}{534.0} \\%& \multirow{3}{*}{Pending} \\ %
     & & AFL & \ntf \\
     & & QC & \ntf  \\
     \midrule
     \multirow{3}{*}{rhino \bugG} & \multirow{3}{*}{\code{IllegalStateException}} &  
     \winner{\tool{}} & \mtf{100}{8.25} \\%& \multirow{3}{*}{Pending} \\ %
     & & AFL & \mtf{20}{5343.0} \\
     & & \winner{QC} & \mtf{100}{9.65}  \\
     \midrule
     \multirow{3}{*}{rhino \bugF} & \multirow{3}{*}{\code{NullPointerException}} & 
     \tool{} & \mtf{100}{18.6} \\%& \multirow{3}{*}{Pending} \\ %
     & & AFL & \ntf \\
     & & \winner{QC} & \mtf{100}{9.85}  \\
     \midrule
     \multirow{3}{*}{rhino \bugH} & \multirow{3}{*}{\code{ClassCastException}} &  
     \winner{\tool{}} & \mtf{85}{245.18} \\%& \multirow{3}{*}{Pending} \\ %
     & & AFL & \ntf \\
     & & QC & \mtf{35}{362.43}  \\
     \midrule
     \multirow{3}{*}{rhino \bugJ} & \multirow{3}{*}{\code{VerifyError}} &  
     \winner{\tool{}} & \mtf{100}{94.75} \\%& \multirow{3}{*}{Confirmed} \\ %
     & & AFL & \ntf \\
     & & QC & \mtf{80}{229.5}  \\
     \midrule
     \multirow{3}{*}{bcel \bugO} & \multirow{3}{*}{\code{ClassFormatException}} &  
     \tool{} & \mtf{100}{19.5} \\%& \multirow{3}{*}{Confirmed} \\ %
     & & \winner{AFL} & \mtf{100}{5.85} \\
     & & QC & \mtf{100}{142.1}  \\
     \midrule
     \multirow{3}{*}{bcel \bugN} & \multirow{3}{*}{\code{AssertionViolatedException}} &  
     \winner{\tool{}} & \mtf{95}{19.32} \\%& \multirow{3}{*}{Pending} \\ %
     & & AFL & \mtf{90}{1082.22} \\
     & & QC & \mtf{5}{15.0}  \\
     \bottomrule
\end{tabular}
\end{adjustbox}
\label{table:bugs}
\end{table*}

\subsection{Bugs in the Semantic Analysis Classes}
\label{sec:eval-bugs}

Each of \tool{}, AFL, and QuickCheck keep track of generated inputs which cause test failures. Ideally, for any given input, the test program should either process it successfully or reject the input as invalid using a documented mechanism, such as throwing a checked  \code{ParseException} on syntax errors. Test \emph{failures} correspond either to assertion violations or to undocumented run-time exceptions being thrown during test execution, such as a \code{NullPointerException}. Test failures can occur during the processing of either valid or invalid inputs; the latter can lead to failures within the syntax or semantic analysis stages themselves.

Across all our experiments, the various fuzzing techniques generated over 95,000 failing inputs that correspond to over 3,000 unique stack traces. We manually triaged these failures by filtering them based on exception type, message text, and source location, resulting in a corpus of what we believe are 20 unique bugs. We have reported each of these bugs to the project developers. At the time of writing: 5 bugs have been fixed, 10 await patches, and 5 reports have received no response.

We classify each bug as \emph{syntactic} or  \emph{semantic}, depending on whether the corresponding exception was raised within the syntactic or semantic analysis classes, respectively (ref. Table~\ref{table:benchmarks}). Of the 20 unique bugs we found, 10 were syntactic and 10 were semantic.

Here, we evaluate \tool{} in discovering \emph{semantic bugs}, for which it is specifically designed. Section~\ref{sec:discussion} discusses the syntactic bugs we found, whose discovery was not \tool{}'s goal. %

Table~\ref{table:bugs} enumerates the 10 semantic bugs that we found across four of the five benchmark programs. The bugs have been given unique IDs---represented as circled letters---for ease of discussion. The table also lists the type of exception thrown for each bug. To evaluate the effectiveness of each of the three techniques in discovering these bugs, we use two metrics. First, we are interested in knowing whether a given technique reliably finds the bug across repeated experiments. We define \emph{reliability} as the percentage of the 20 runs (of 3-hours each) in which a given technique finds a particular bug at least once. Second, we measure the \emph{mean time to find} (MTF) the first input that triggers the given bug, across the repetitions in which it was found. Naturally, a shorter MTF is desirable. For each bug, we circle the name of the technique that is the most effective in finding that bug. We define \emph{most effective} as the technique with either the highest reliability, or if there is a tie in reliability, then the shortest MTF. 

The table indicates that \tool{} is the most effective technique in finding 8 of the 10 bugs; in the remaining two cases (\bugF{} and \bugO{}), \tool{} still finds the bugs with 100\% reliability and in less than 20 seconds on average. In fact, \tool{} finds all the 10 semantic bugs in \emph{at most 10 minutes on average}; 7 are found within the first 2 minutes on average. In contrast, AFL requires more than one hour to find 3 of the bugs (\bugB{}, \bugC{}, \bugG{}), and simply does not find 5 of the bugs within the 3-hour time limit. This makes sense because AFL's mutations on the raw input strings do not guarantee syntactic validity; it generates much fewer inputs that reach the semantic analysis stage. QuickCheck discovers 8 of the 10 semantic bugs, but since it relies on random sampling alone, its reliability is often low. For example, QuickCheck discovers \bugB{} only 10\% of the time, and \bugN{} only 5\% of the time; Zest discovers them 100\% and 95\% of the time, respectively. Overall, \tool{} is clearly the most effective technique in discovering bugs in the semantic analysis classes of our benchmarks.

\paragraph{Case studies} %

In Ant, \bugB{} is triggered when the input \code{build.xml} document contains both an \code{<augment>} element and a \code{<target>} element inside the root \code{<project>} element, but when the \code{<augment>} element is missing an \code{id} attribute. This incomplete semantic check leads to an \code{IllegalStateException} for a component down the pipeline which tries to configure an Ant task. Following our bug report, this issue has been fixed starting Ant version 1.10.6.

In Rhino, \bugJ{} is triggered by a semantically valid input. Rhino successfully validates the input JavaScript program and then compiles it to Java bytecode. However, the compiled bytecode is corrupted, which results in a \code{VerifyError} being generated by the JVM. AFL does not find this bug at all. The Rhino developers confirmed the bug, though a fix is still pending.

In Closure, \bugC{} is an NPE that is triggered in its dead-code elimination pass when handling arrow functions that reference undeclared variables, such as \code{"x => y"}. The generator-based techniques always find this bug and within just 8.8 seconds on average, while AFL requires more than 90 minutes and only finds it in 20\% of the runs. The Closure developers fixed this issue after our report.

\bugD{} is a bug in Closure's semantic analysis of variable declarations. The bug is triggered when a new variable is declared after a \code{break} statement. Although everything immediately after a \code{break} statement is unreachable code, variable declarations in JavaScript are hoisted and therefore cannot be removed. \tool{} is the only technique that discovered this bug. A sample input \tool{} generated is:

\begin{center}
\small
\begin{verbatim}
while ((l_0)){ 
  while ((l_0)){ 
    if ((l_0)) { break;;var l_0;continue }
    { break;var l_0 } 
  } 
}
\end{verbatim}
\end{center}

\bugU{} was the most elusive bug that we encountered. \tool{} is the only technique that finds it and it does so in only one of the 20 runs. An exception is triggered by the following input:

\begin{center}
\small
\begin{verbatim}
((o_0) => (((o_0) *= (o_0)) 
   < ((i_1) &= ((o_0)((((undefined)[(((i_1, o_0, a_2) => { 
      if ((i_1)) { throw ((false).o_0) } 
  })((y_3)))])((new (null)((true))))))))))
\end{verbatim}
\end{center}

\noindent The issue is perhaps rooted in Closure's attempt to compile-time evaluate \code{undefined[undefined](...)}. The developers acknowledged the bug but have not yet published a fix. These complex examples demonstrate both the power of \tool{}'s generators, which reduce the search space to syntactically valid inputs, as well as the effectiveness of its feedback-directed parameter search. %

\section{Discussion and Threats to Validity}
\label{sec:discussion}

\tool{} and QuickCheck make use of generators for synthesizing inputs that are syntactically valid by construction. By design, these tools do not exercise code paths corresponding to parse errors in the {syntax analysis} stage. In contrast, AFL performs mutations directly on raw input strings. Byte-level mutations on raw inputs usually lead to inputs that do not parse. Consequently, AFL spends most of its time testing error paths within the syntax analysis stages.

In our experiments, AFL achieved the highest coverage within the syntax analysis classes of our benchmarks (ref. Table~\ref{table:benchmarks}), $1.1\times$-$1.6\times$ higher than \tool{}'s syntax analysis coverage. 
Further, AFL discovered 10 syntactic bugs in addition to the bugs enumerated in Table~\ref{table:benchmarks}: 3 in Maven, 6 in BCEL, and 1 in Rhino. These bugs were triggered  by syntactically invalid inputs, which the generator-based tools do not produce. \tool{} does not attempt to target these bugs; rather, it is complementary to AFL-like tools.

\tool{} assumes the availability of QuickCheck-like generators to exercise the semantic analysis classes and to find semantic bugs. Although this is no doubt an additional cost, the effort required to develop a structured-input generator is usually no more than the effort required to write unit tests with hand-crafted structured inputs, which is usually an accepted cost. In fact, due to the growing popularity of generator-based testing tools like Hypothesis~\cite{hypothesis}, ScalaCheck~\cite{scalacheck}, PropEr~\cite{Papadakis11}, etc. a large number of off-the-shelf or automatically synthesized type-based generators are available. The \tool{} technique can, in principle, work with any such generator. When given a generator, \tool{} excels at exercising semantic analyses and is very effective in discovering semantic bugs.%

We did not evaluate how \tool{}'s effectiveness might vary depending on the quality of generators, since we hand-wrote the simplest generators possible for our benchmarks. 
However,  our results suggest that \tool{}'s ability to guide generation towards paths deep in the semantic analysis stage make its performance less tied to generator quality than pure random sampling as in QuickCheck.

The effectiveness of \cgf{} tools like AFL is usually sensitive to the choice of seed inputs~\cite{Klees18}. Although the relative differences between the performance of \tool{} and AFL will likely vary with this choice, the purpose of our evaluation was to demonstrate that focusing on feedback-directed search in the space of syntactically valid inputs is advantageous. No matter what seed inputs one provides to conventional fuzzing tools, the byte-level mutations on raw inputs will lead to an enormous number of syntax errors. We believe that approaches like \tool{} complement \cgf{} tools in testing different components of programs.

\vspace{-0.1cm}
\section{Related Work}
\label{sec:related}

A lot of research has gone into automated test-input generation techniques, as surveyed by Anand et al.~\cite{AnandBCCCGHHMOE13}.

Randoop~\cite{Pacheco07} and EvoSuite~\cite{Fraser11} generate JUnit tests for a particular class by incrementally trying and combining sequences of calls.  During the generation of sequence of calls, both Randoop and EvoSuite take some form of feedback into account.  %
These tools produce unit tests by directly invoking methods on the component classes. In contrast, \tool{} addresses the problem of generating \emph{inputs} when given a test driver and an input generator.

UDITA~\cite{Gligoric10} allows developers to write random-input generators in a QuickCheck-like language. UDITA then performs \emph{bounded-exhaustive} enumeration of the paths through the generators, along with several optimizations. In contrast, \tool{} relies on random mutations over the entire parameter space but uses code coverage and input-validity feedback to guide the parameter search. It would be interesting to see if UDITA's search strategy could be augmented to selectively adjust bounds using code coverage and validity feedback; however, we leave this investigation as future work. 

Targeted property-testing~\cite{Loscher17, Loscher18} guides input generators used in property testing towards a user-specified fitness value using techniques such as hill climbing and simulated annealing. G\"odelTest~\cite{Feldt13} attempts to satisfy user-specified properties on inputs. It performs a meta-heuristic search for stochastic models that are used to sample random inputs from a generator. Unlike these techniques, \tool{} relies on code coverage feedback to guide input generation towards exploring diverse program behaviors.

In the security community, several tools have been developed to improve the effectiveness of coverage-guided fuzzing in reaching deep program states~\cite{Rawat17, Li17, Bohme17, Lemieux18-FairFuzz, Chen18}.
AFLGo~\cite{Bohme17} extends AFL to direct fuzzing towards generating inputs that exercise a program point of interest.
It relies on call graphs obtained from whole-program static analysis, which can be difficult to compute precisely in our ecosystem. \tool{} is purely a dynamic analysis tool.
FairFuzz~\cite{Lemieux18-FairFuzz} modifies AFL to bias input generation towards branches that are rarely executed, but does not explicitly identify parts of the program that perform the core logic. In \tool{}, we bias input generation towards validity no matter how frequently the semantic analysis stage is exercised.%

\tool{} generates inputs that are syntactically valid by construction (assuming suitable generators), but uses heuristics to guide input generation towards semantic validity. Unlike \tool{}, symbolic execution tools~\cite{Clarke76, King76, Godefroid05,Sen05,Cadar08,Chipounov12, Li11,Tillmann08,Anand07,Godefroid08,Avgerinos14} methodically explore the program under test by capturing path constraints and can directly generate inputs which satisfy specific path constraints. Symbolic execution can thus precisely produce valid inputs exercising new behavior. The cost of this precision is that it can lead to the path explosion problem for larger programs. Hybrid techniques that combine symbolic execution with coverage-guided fuzzing have been proposed~\cite{Bottinger16, Stephens16, Ognawala18, Yun18}. These hybrid techniques could be combined with \tool{} to solve for parameter sequences that satisfy branch constraints which \tool{} may not cover on its own.

Grammar-based fuzzing~\cite{Maurer90, Coppit05, Sirer99, Godefroid08, BeyeneA12} techniques rely on context-free grammar specifications to generate structured inputs. %
CSmith~\cite{Yang11} generates random C programs for differential testing of C compilers. LangFuzz~\cite{Holler12} generates random programs using a grammar and by recombining code fragments from a codebase. These approaches fall under the category of generator-based testing, but primarily focus on tuning the underlying generators rather than using code coverage feedback. \tool{} is not restricted to context-free grammars, and does not require any domain-specific tuning. 
 
Several recently developed tools leverage input format specifications (either  grammmars~\cite{Aschermann19,Wang19}, file formats~\cite{Pham18}, or protocol buffers~\cite{Serebryany17}) to improve the performance of \cgf{}. These tools develop mutations that are specific to the input format specifications, e.g. syntax-tree mutations for grammars. \tool{}'s generators are arbitrary programs; therefore, we perform mutations directly on the parameters that determine the execution path through the generators,
rather than on a parsed representation of inputs.

There has also been some recent interest in automatically generating input grammars from existing inputs, using machine learning~\cite{Godefroid17} and language inference algorithms \cite{Bastani17}. Similarly, DIFUZE~\cite{Corina17} infers device driver interfaces from a running kernel to boostrap subsequent structured fuzzing. These techniques are complementary to \tool{}---the grammars generated by these techniques could be transformed into parametric generators for \tool{}. 

Finally, validity feedback has been useful in fuzzing digital circuits that have constrained interfaces~\cite{Laeufer18}, as well as in generating seed inputs for conventional fuzzing~\cite{Wang17}.

\section{Conclusion}

We presented \tool{}, a technique that incorporates ideas from coverage-guided fuzzing into generator-based testing. We showed how a simple conversion of random-input generators into \emph{parametric} input generators enables an elegant mapping from low-level mutations in the untyped parameter domain into structural mutations in the syntactically valid input domain. We then presented an algorithm that combined code coverage feedback with input-validity feedback to guide the generation of test inputs. On 5 real-world Java benchmarks, we found that \tool{} achieved consistently higher branch coverage and had better bug-finding ability in the semantic analysis stage than baseline techniques. Our results suggest that \tool{} is highly effective at testing the semantic analysis stage of programs, complementing tools such as AFL that are effective at testing the syntactic analysis stage of programs.

\begin{acks}
We would like to thank Kevin Laeufer and Michael Dennis for their helpful feedback during the development of this project, as well as Benjamin Brock and Rohan Bavishi for their invaluable comments on the paper draft. 
This research is supported in part by gifts from Samsung, Facebook, and Futurewei, and by NSF grants CCF-1409872 and CNS-1817122.%
\end{acks}

\break
\balance
\bibliographystyle{ACM-Reference-Format}
\bibliography{zest}


\begin{thebibliography}{75}

%%% the Web address.

\ifx \showCODEN    \undefined \def \showCODEN     #1{\unskip}     \fi
\ifx \showDOI      \undefined \def \showDOI       #1{#1}\fi
\ifx \showISBNx    \undefined \def \showISBNx     #1{\unskip}     \fi
\ifx \showISBNxiii \undefined \def \showISBNxiii  #1{\unskip}     \fi
\ifx \showISSN     \undefined \def \showISSN      #1{\unskip}     \fi
\ifx \showLCCN     \undefined \def \showLCCN      #1{\unskip}     \fi
\ifx \shownote     \undefined \def \shownote      #1{#1}          \fi
\ifx \showarticletitle \undefined \def \showarticletitle #1{#1}   \fi
\ifx \showURL      \undefined \def \showURL       {\relax}        \fi
% invisible to TeX
\providecommand\bibfield[2]{#2}
\providecommand\bibinfo[2]{#2}
\providecommand\natexlab[1]{#1}
\providecommand\showeprint[2][]{arXiv:#2}

\bibitem[\protect\citeauthoryear{??}{ant}{2018}]%
        {ant}
 \bibinfo{year}{2018}\natexlab{}.
\newblock \bibinfo{title}{{Apache Ant}}.
\newblock \bibinfo{howpublished}{\url{https://ant.apache.org}}.
\newblock
\newblock
\shownote{Accessed August 24, 2018.}


\bibitem[\protect\citeauthoryear{??}{bce}{2018}]%
        {bcel}
 \bibinfo{year}{2018}\natexlab{}.
\newblock \bibinfo{title}{{Apache Byte Code Engineering Library}}.
\newblock
  \bibinfo{howpublished}{\url{https://commons.apache.org/proper/commons-bcel}}.
\newblock
\newblock
\shownote{Accessed August 24, 2018.}


\bibitem[\protect\citeauthoryear{??}{mav}{2018}]%
        {maven}
 \bibinfo{year}{2018}\natexlab{}.
\newblock \bibinfo{title}{{Apache Maven}}.
\newblock \bibinfo{howpublished}{\url{https://maven.apache.org}}.
\newblock
\newblock
\shownote{Accessed August 24, 2018.}


\bibitem[\protect\citeauthoryear{??}{clo}{2018}]%
        {closure}
 \bibinfo{year}{2018}\natexlab{}.
\newblock \bibinfo{title}{{Google Closure}}.
\newblock
  \bibinfo{howpublished}{\url{https://developers.google.com/closure/compiler}}.
\newblock
\newblock
\shownote{Accessed August 24, 2018.}


\bibitem[\protect\citeauthoryear{??}{rhi}{2018}]%
        {rhino}
 \bibinfo{year}{2018}\natexlab{}.
\newblock \bibinfo{title}{{Mozilla Rhino}}.
\newblock \bibinfo{howpublished}{\url{https://github.com/mozilla/rhino}}.
\newblock
\newblock
\shownote{Accessed August 24, 2018.}


\bibitem[\protect\citeauthoryear{??}{rea}{2018}]%
        {react}
 \bibinfo{year}{2018}\natexlab{}.
\newblock \bibinfo{title}{{React.JS}}.
\newblock \bibinfo{howpublished}{\url{https://reactjs.org}}.
\newblock
\newblock
\shownote{Accessed August 24, 2018.}


\bibitem[\protect\citeauthoryear{??}{bes}{2019}]%
        {bestorm}
 \bibinfo{year}{2019}\natexlab{}.
\newblock \bibinfo{title}{{beStorm\textregistered Software Security}}.
\newblock
  \bibinfo{howpublished}{\url{https://www.beyondsecurity.com/bestorm.html}}.
\newblock
\newblock
\shownote{Accessed January 28, 2019.}


\bibitem[\protect\citeauthoryear{??}{cod}{2019}]%
        {codenomicon}
 \bibinfo{year}{2019}\natexlab{}.
\newblock \bibinfo{title}{{Codenomicon Vulnerability Management}}.
\newblock \bibinfo{howpublished}{\url{http://www.codenomicon.com/index.html}}.
\newblock
\newblock
\shownote{Accessed January 28, 2019.}


\bibitem[\protect\citeauthoryear{??}{cyb}{2019}]%
        {cyberflood}
 \bibinfo{year}{2019}\natexlab{}.
\newblock \bibinfo{title}{{CyberFlood - Spirent}}.
\newblock
  \bibinfo{howpublished}{\url{https://www.spirent.com/products/cyberflood}}.
\newblock
\newblock
\shownote{Accessed January 28, 2019.}


\bibitem[\protect\citeauthoryear{??}{eri}{2019}]%
        {eris}
 \bibinfo{year}{2019}\natexlab{}.
\newblock \bibinfo{title}{{Eris: Porting of QuickCheck to PHP}}.
\newblock \bibinfo{howpublished}{\url{https://github.com/giorgiosironi/eris}}.
\newblock
\newblock
\shownote{Accessed January 28, 2019.}


\bibitem[\protect\citeauthoryear{??}{hyp}{2019}]%
        {hypothesis}
 \bibinfo{year}{2019}\natexlab{}.
\newblock \bibinfo{title}{{Hypothesis for Python}}.
\newblock \bibinfo{howpublished}{\url{https://hypothesis.works/}}.
\newblock
\newblock
\shownote{Accessed January 28, 2019.}


\bibitem[\protect\citeauthoryear{??}{jsv}{2019}]%
        {jsverify}
 \bibinfo{year}{2019}\natexlab{}.
\newblock \bibinfo{title}{{JSVerify: Property-based testing for JavaScript}}.
\newblock \bibinfo{howpublished}{\url{https://github.com/jsverify/jsverify}}.
\newblock
\newblock
\shownote{Accessed January 28, 2019.}


\bibitem[\protect\citeauthoryear{??}{pea}{2019}]%
        {peach}
 \bibinfo{year}{2019}\natexlab{}.
\newblock \bibinfo{title}{{PeachFuzzer}}.
\newblock \bibinfo{howpublished}{\url{https://www.peach.tech/}}.
\newblock
\newblock
\shownote{Accessed January 28, 2019.}


\bibitem[\protect\citeauthoryear{??}{sca}{2019}]%
        {scalacheck}
 \bibinfo{year}{2019}\natexlab{}.
\newblock \bibinfo{title}{{ScalaCheck: Property-based testing for Scala}}.
\newblock \bibinfo{howpublished}{\url{https://www.scalacheck.org/}}.
\newblock
\newblock
\shownote{Accessed January 28, 2019.}


\bibitem[\protect\citeauthoryear{??}{tes}{2019}]%
        {test.check}
 \bibinfo{year}{2019}\natexlab{}.
\newblock \bibinfo{title}{{test.check: QuickCheck for Clojure}}.
\newblock \bibinfo{howpublished}{\url{https://github.com/clojure/test.check}}.
\newblock
\newblock
\shownote{Accessed January 28, 2019.}


\bibitem[\protect\citeauthoryear{Amaral, Florido, and Costa}{Amaral
  et~al\mbox{.}}{2014}]%
        {Amaral14}
\bibfield{author}{\bibinfo{person}{Cl{\'a}udio Amaral},
  \bibinfo{person}{M{\'a}rio Florido}, {and} \bibinfo{person}{V{\'\i}tor~Santos
  Costa}.} \bibinfo{year}{2014}\natexlab{}.
\newblock \showarticletitle{PrologCheck--property-based testing in Prolog}. In
  \bibinfo{booktitle}{\emph{International Symposium on Functional and Logic
  Programming}}. Springer, \bibinfo{pages}{1--17}.
\newblock


\bibitem[\protect\citeauthoryear{Anand, Burke, Chen, Clark, Cohen, Grieskamp,
  Harman, Harrold, and McMinn}{Anand et~al\mbox{.}}{2013}]%
        {AnandBCCCGHHMOE13}
\bibfield{author}{\bibinfo{person}{Saswat Anand}, \bibinfo{person}{Edmund~K.
  Burke}, \bibinfo{person}{Tsong~Yueh Chen}, \bibinfo{person}{John~A. Clark},
  \bibinfo{person}{Myra~B. Cohen}, \bibinfo{person}{Wolfgang Grieskamp},
  \bibinfo{person}{Mark Harman}, \bibinfo{person}{Mary~Jean Harrold}, {and}
  \bibinfo{person}{Phil McMinn}.} \bibinfo{year}{2013}\natexlab{}.
\newblock \showarticletitle{An orchestrated survey of methodologies for
  automated software test case generation}.
\newblock \bibinfo{journal}{\emph{Journal of Systems and Software}}
  \bibinfo{volume}{86}, \bibinfo{number}{8} (\bibinfo{year}{2013}),
  \bibinfo{pages}{1978--2001}.
\newblock
\urldef\tempurl%
\url{https://doi.org/10.1016/j.jss.2013.02.061}
\showDOI{\tempurl}


\bibitem[\protect\citeauthoryear{Anand, P\u{a}s\u{a}reanu, and Visser}{Anand
  et~al\mbox{.}}{2007}]%
        {Anand07}
\bibfield{author}{\bibinfo{person}{Saswat Anand}, \bibinfo{person}{Corina~S.
  P\u{a}s\u{a}reanu}, {and} \bibinfo{person}{Willem Visser}.}
  \bibinfo{year}{2007}\natexlab{}.
\newblock \showarticletitle{{JPF-SE}: a symbolic execution extension to {Java
  PathFinder}}. In \bibinfo{booktitle}{\emph{Proceedings of Tools and
  Algorithms for the Construction and Analysis of Systems (TACAS)}}.
\newblock


\bibitem[\protect\citeauthoryear{Arts, Hughes, Johansson, and Wiger}{Arts
  et~al\mbox{.}}{2006}]%
        {Arts06}
\bibfield{author}{\bibinfo{person}{Thomas Arts}, \bibinfo{person}{John Hughes},
  \bibinfo{person}{Joakim Johansson}, {and} \bibinfo{person}{Ulf~T. Wiger}.}
  \bibinfo{year}{2006}\natexlab{}.
\newblock \showarticletitle{Testing telecoms software with quviq QuickCheck}.
  In \bibinfo{booktitle}{\emph{Proceedings of the 2006 {ACM} {SIGPLAN} Workshop
  on Erlang, Portland, Oregon, USA, September 16, 2006}}.
  \bibinfo{pages}{2--10}.
\newblock
\urldef\tempurl%
\url{https://doi.org/10.1145/1159789.1159792}
\showDOI{\tempurl}


\bibitem[\protect\citeauthoryear{Aschermann, Frassetto, Holz, Jauernig,
  Sadeghi, and Teuchert}{Aschermann et~al\mbox{.}}{2019}]%
        {Aschermann19}
\bibfield{author}{\bibinfo{person}{Cornelius Aschermann},
  \bibinfo{person}{Tommaso Frassetto}, \bibinfo{person}{Thorsten Holz},
  \bibinfo{person}{Patrick Jauernig}, \bibinfo{person}{Ahmad-Reza Sadeghi},
  {and} \bibinfo{person}{Daniel Teuchert}.} \bibinfo{year}{2019}\natexlab{}.
\newblock \showarticletitle{{Nautilus: Fishing for Deep Bugs with Grammars}}.
  In \bibinfo{booktitle}{\emph{26th Annual Network and Distributed System
  Security Symposium}} \emph{(\bibinfo{series}{NDSS '19})}.
\newblock


\bibitem[\protect\citeauthoryear{Avgerinos, Rebert, Cha, and Brumley}{Avgerinos
  et~al\mbox{.}}{2014}]%
        {Avgerinos14}
\bibfield{author}{\bibinfo{person}{Thanassis Avgerinos},
  \bibinfo{person}{Alexandre Rebert}, \bibinfo{person}{Sang~Kil Cha}, {and}
  \bibinfo{person}{David Brumley}.} \bibinfo{year}{2014}\natexlab{}.
\newblock \showarticletitle{Enhancing Symbolic Execution with Veritesting}. In
  \bibinfo{booktitle}{\emph{Proceedings of the 36th International Conference on
  Software Engineering}} \emph{(\bibinfo{series}{ICSE 2014})}.
  \bibinfo{publisher}{ACM}, \bibinfo{address}{New York, NY, USA},
  \bibinfo{pages}{1083--1094}.
\newblock
\showISBNx{978-1-4503-2756-5}


\bibitem[\protect\citeauthoryear{Bastani, Sharma, Aiken, and Liang}{Bastani
  et~al\mbox{.}}{2017}]%
        {Bastani17}
\bibfield{author}{\bibinfo{person}{Osbert Bastani}, \bibinfo{person}{Rahul
  Sharma}, \bibinfo{person}{Alex Aiken}, {and} \bibinfo{person}{Percy Liang}.}
  \bibinfo{year}{2017}\natexlab{}.
\newblock \showarticletitle{Synthesizing Program Input Grammars}. In
  \bibinfo{booktitle}{\emph{Proceedings of the 38th ACM SIGPLAN Conference on
  Programming Language Design and Implementation}} \emph{(\bibinfo{series}{PLDI
  2017})}. \bibinfo{publisher}{ACM}, \bibinfo{address}{New York, NY, USA},
  \bibinfo{pages}{95--110}.
\newblock
\showISBNx{978-1-4503-4988-8}
\urldef\tempurl%
\url{https://doi.org/10.1145/3062341.3062349}
\showDOI{\tempurl}


\bibitem[\protect\citeauthoryear{Beyene and Andrews}{Beyene and
  Andrews}{2012}]%
        {BeyeneA12}
\bibfield{author}{\bibinfo{person}{Michael Beyene} {and}
  \bibinfo{person}{James~H. Andrews}.} \bibinfo{year}{2012}\natexlab{}.
\newblock \showarticletitle{Generating String Test Data for Code Coverage}. In
  \bibinfo{booktitle}{\emph{Fifth {IEEE} International Conference on Software
  Testing, Verification and Validation, {ICST} 2012, Montreal, QC, Canada,
  April 17-21, 2012}}. \bibinfo{pages}{270--279}.
\newblock
\urldef\tempurl%
\url{https://doi.org/10.1109/ICST.2012.107}
\showDOI{\tempurl}


\bibitem[\protect\citeauthoryear{B\"{o}hme, Pham, Nguyen, and
  Roychoudhury}{B\"{o}hme et~al\mbox{.}}{2017}]%
        {Bohme17}
\bibfield{author}{\bibinfo{person}{Marcel B\"{o}hme},
  \bibinfo{person}{Van-Thuan Pham}, \bibinfo{person}{Manh-Dung Nguyen}, {and}
  \bibinfo{person}{Abhik Roychoudhury}.} \bibinfo{year}{2017}\natexlab{}.
\newblock \showarticletitle{Directed Greybox Fuzzing}. In
  \bibinfo{booktitle}{\emph{Proceedings of the 2017 ACM SIGSAC Conference on
  Computer and Communications Security}} \emph{(\bibinfo{series}{CCS '17})}.
\newblock


\bibitem[\protect\citeauthoryear{B\"{o}hme, Pham, and Roychoudhury}{B\"{o}hme
  et~al\mbox{.}}{2016}]%
        {Bohme16}
\bibfield{author}{\bibinfo{person}{Marcel B\"{o}hme},
  \bibinfo{person}{Van-Thuan Pham}, {and} \bibinfo{person}{Abhik
  Roychoudhury}.} \bibinfo{year}{2016}\natexlab{}.
\newblock \showarticletitle{Coverage-based Greybox Fuzzing As Markov Chain}. In
  \bibinfo{booktitle}{\emph{Proceedings of the 2016 ACM SIGSAC Conference on
  Computer and Communications Security}} \emph{(\bibinfo{series}{CCS '16})}.
\newblock


\bibitem[\protect\citeauthoryear{B\"{o}ttinger and Eckert}{B\"{o}ttinger and
  Eckert}{2016}]%
        {Bottinger16}
\bibfield{author}{\bibinfo{person}{Konstantin B\"{o}ttinger} {and}
  \bibinfo{person}{Claudia Eckert}.} \bibinfo{year}{2016}\natexlab{}.
\newblock \showarticletitle{{DeepFuzz}: Triggering Vulnerabilities Deeply
  Hidden in Binaries}. In \bibinfo{booktitle}{\emph{Proceedings of the 13th
  International Conference on Detection of Intrusions and Malware, and
  Vulnerability Assessment - Volume 9721}} \emph{(\bibinfo{series}{DIMVA
  2016})}. \bibinfo{publisher}{Springer-Verlag}, \bibinfo{address}{Berlin,
  Heidelberg}, \bibinfo{pages}{25--34}.
\newblock
\showISBNx{978-3-319-40666-4}
\urldef\tempurl%
\url{https://doi.org/10.1007/978-3-319-40667-1_2}
\showDOI{\tempurl}


\bibitem[\protect\citeauthoryear{Cadar, Dunbar, and Engler}{Cadar
  et~al\mbox{.}}{2008}]%
        {Cadar08}
\bibfield{author}{\bibinfo{person}{Cristian Cadar}, \bibinfo{person}{Daniel
  Dunbar}, {and} \bibinfo{person}{Dawson Engler}.}
  \bibinfo{year}{2008}\natexlab{}.
\newblock \showarticletitle{{KLEE: Unassisted and Automatic Generation of
  High-coverage Tests for Complex Systems Programs}}. In
  \bibinfo{booktitle}{\emph{Proceedings of the 8th USENIX Conference on
  Operating Systems Design and Implementation}}
  \emph{(\bibinfo{series}{OSDI'08})}.
\newblock


\bibitem[\protect\citeauthoryear{Chen and Chen}{Chen and Chen}{2018}]%
        {Chen18}
\bibfield{author}{\bibinfo{person}{Peng Chen} {and} \bibinfo{person}{Hao
  Chen}.} \bibinfo{year}{2018}\natexlab{}.
\newblock \showarticletitle{{Angora: Efficient Fuzzing by Principled Search}}.
  In \bibinfo{booktitle}{\emph{Proceedings of the 39th IEEE Symposium on
  Security and Privacy}}.
\newblock


\bibitem[\protect\citeauthoryear{Chipounov, Kuznetsov, and Candea}{Chipounov
  et~al\mbox{.}}{2012}]%
        {Chipounov12}
\bibfield{author}{\bibinfo{person}{Vitaly Chipounov},
  \bibinfo{person}{Volodymyr Kuznetsov}, {and} \bibinfo{person}{George
  Candea}.} \bibinfo{year}{2012}\natexlab{}.
\newblock \showarticletitle{The S2E Platform: Design, Implementation, and
  Applications}.
\newblock \bibinfo{journal}{\emph{ACM Transactions on Computer Systems.}}
  \bibinfo{volume}{30}, \bibinfo{number}{1} (\bibinfo{year}{2012}),
  \bibinfo{pages}{2}.
\newblock


\bibitem[\protect\citeauthoryear{Claessen and Hughes}{Claessen and
  Hughes}{2000}]%
        {Claessen00}
\bibfield{author}{\bibinfo{person}{Koen Claessen} {and} \bibinfo{person}{John
  Hughes}.} \bibinfo{year}{2000}\natexlab{}.
\newblock \showarticletitle{QuickCheck: A Lightweight Tool for Random Testing
  of Haskell Programs}. In \bibinfo{booktitle}{\emph{Proceedings of the 5th ACM
  SIGPLAN International Conference on Functional Programming}}
  \emph{(\bibinfo{series}{ICFP})}.
\newblock


\bibitem[\protect\citeauthoryear{Clarke}{Clarke}{1976}]%
        {Clarke76}
\bibfield{author}{\bibinfo{person}{Lori~A. Clarke}.}
  \bibinfo{year}{1976}\natexlab{}.
\newblock \showarticletitle{A program testing system}. In
  \bibinfo{booktitle}{\emph{Proc. of the 1976 annual conference}}.
  \bibinfo{pages}{488--491}.
\newblock


\bibitem[\protect\citeauthoryear{Coppit and Lian}{Coppit and Lian}{2005}]%
        {Coppit05}
\bibfield{author}{\bibinfo{person}{David Coppit} {and} \bibinfo{person}{Jiexin
  Lian}.} \bibinfo{year}{2005}\natexlab{}.
\newblock \showarticletitle{Yagg: An Easy-to-use Generator for Structured Test
  Inputs}. In \bibinfo{booktitle}{\emph{Proceedings of the 20th IEEE/ACM
  International Conference on Automated Software Engineering}}
  \emph{(\bibinfo{series}{ASE '05})}. \bibinfo{publisher}{ACM},
  \bibinfo{address}{New York, NY, USA}, \bibinfo{pages}{356--359}.
\newblock
\showISBNx{1-58113-993-4}
\urldef\tempurl%
\url{https://doi.org/10.1145/1101908.1101969}
\showDOI{\tempurl}


\bibitem[\protect\citeauthoryear{Corina, Machiry, Salls, Shoshitaishvili, Hao,
  Kruegel, and Vigna}{Corina et~al\mbox{.}}{2017}]%
        {Corina17}
\bibfield{author}{\bibinfo{person}{Jake Corina}, \bibinfo{person}{Aravind
  Machiry}, \bibinfo{person}{Christopher Salls}, \bibinfo{person}{Yan
  Shoshitaishvili}, \bibinfo{person}{Shuang Hao}, \bibinfo{person}{Christopher
  Kruegel}, {and} \bibinfo{person}{Giovanni Vigna}.}
  \bibinfo{year}{2017}\natexlab{}.
\newblock \showarticletitle{{DIFUZE}: Interface Aware Fuzzing for Kernel
  Drivers}. In \bibinfo{booktitle}{\emph{Proceedings of the 2017 ACM SIGSAC
  Conference on Computer and Communications Security}}
  \emph{(\bibinfo{series}{CCS '17})}. \bibinfo{publisher}{ACM},
  \bibinfo{address}{New York, NY, USA}, \bibinfo{pages}{2123--2138}.
\newblock
\showISBNx{978-1-4503-4946-8}
\urldef\tempurl%
\url{https://doi.org/10.1145/3133956.3134069}
\showDOI{\tempurl}


\bibitem[\protect\citeauthoryear{Emek, Jaeger, Naveh, Bergman, Aloni, Katz,
  Farkash, Dozoretz, and Goldin}{Emek et~al\mbox{.}}{2002}]%
        {Emek02}
\bibfield{author}{\bibinfo{person}{Roy Emek}, \bibinfo{person}{Itai Jaeger},
  \bibinfo{person}{Yehuda Naveh}, \bibinfo{person}{Gadi Bergman},
  \bibinfo{person}{Guy Aloni}, \bibinfo{person}{Yoav Katz},
  \bibinfo{person}{Monica Farkash}, \bibinfo{person}{Igor Dozoretz}, {and}
  \bibinfo{person}{Alex Goldin}.} \bibinfo{year}{2002}\natexlab{}.
\newblock \showarticletitle{X-Gen: A random test-case generator for systems and
  SoCs}. In \bibinfo{booktitle}{\emph{High-Level Design Validation and Test
  Workshop, 2002. Seventh IEEE International}}. IEEE,
  \bibinfo{pages}{145--150}.
\newblock


\bibitem[\protect\citeauthoryear{Feldt and Poulding}{Feldt and
  Poulding}{2013}]%
        {Feldt13}
\bibfield{author}{\bibinfo{person}{Robert Feldt} {and} \bibinfo{person}{Simon
  Poulding}.} \bibinfo{year}{2013}\natexlab{}.
\newblock \showarticletitle{Finding test data with specific properties via
  metaheuristic search}. In \bibinfo{booktitle}{\emph{2013 IEEE 24th
  International Symposium on Software Reliability Engineering (ISSRE)}}. IEEE,
  \bibinfo{pages}{350--359}.
\newblock


\bibitem[\protect\citeauthoryear{Fraser and Arcuri}{Fraser and Arcuri}{2011}]%
        {Fraser11}
\bibfield{author}{\bibinfo{person}{Gordon Fraser} {and} \bibinfo{person}{Andrea
  Arcuri}.} \bibinfo{year}{2011}\natexlab{}.
\newblock \showarticletitle{EvoSuite: Automatic Test Suite Generation for
  Object-oriented Software}. In \bibinfo{booktitle}{\emph{Proceedings of the
  19th ACM SIGSOFT Symposium and the 13th European Conference on Foundations of
  Software Engineering}} \emph{(\bibinfo{series}{ESEC/FSE '11})}.
\newblock


\bibitem[\protect\citeauthoryear{Fraser and Arcuri}{Fraser and Arcuri}{2014}]%
        {Fraser14}
\bibfield{author}{\bibinfo{person}{Gordon Fraser} {and} \bibinfo{person}{Andrea
  Arcuri}.} \bibinfo{year}{2014}\natexlab{}.
\newblock \showarticletitle{A Large-Scale Evaluation of Automated Unit Test
  Generation Using EvoSuite}.
\newblock \bibinfo{journal}{\emph{ACM Trans. Softw. Eng. Methodol.}}
  \bibinfo{volume}{24}, \bibinfo{number}{2}, Article \bibinfo{articleno}{8}
  (\bibinfo{date}{Dec.} \bibinfo{year}{2014}), \bibinfo{numpages}{42}~pages.
\newblock
\showISSN{1049-331X}
\urldef\tempurl%
\url{https://doi.org/10.1145/2685612}
\showDOI{\tempurl}


\bibitem[\protect\citeauthoryear{Gligoric, Gvero, Jagannath, Khurshid, Kuncak,
  and Marinov}{Gligoric et~al\mbox{.}}{2010}]%
        {Gligoric10}
\bibfield{author}{\bibinfo{person}{Milos Gligoric}, \bibinfo{person}{Tihomir
  Gvero}, \bibinfo{person}{Vilas Jagannath}, \bibinfo{person}{Sarfraz
  Khurshid}, \bibinfo{person}{Viktor Kuncak}, {and} \bibinfo{person}{Darko
  Marinov}.} \bibinfo{year}{2010}\natexlab{}.
\newblock \showarticletitle{Test generation through programming in {UDITA}}. In
  \bibinfo{booktitle}{\emph{Proceedings of the 32nd {ACM/IEEE} International
  Conference on Software Engineering - Volume 1, {ICSE} 2010, Cape Town, South
  Africa, 1-8 May 2010}}. \bibinfo{pages}{225--234}.
\newblock
\urldef\tempurl%
\url{https://doi.org/10.1145/1806799.1806835}
\showDOI{\tempurl}


\bibitem[\protect\citeauthoryear{Godefroid, Kiezun, and Levin}{Godefroid
  et~al\mbox{.}}{2008}]%
        {Godefroid08}
\bibfield{author}{\bibinfo{person}{Patrice Godefroid}, \bibinfo{person}{Adam
  Kiezun}, {and} \bibinfo{person}{Michael~Y. Levin}.}
  \bibinfo{year}{2008}\natexlab{}.
\newblock \showarticletitle{Grammar-based Whitebox Fuzzing}. In
  \bibinfo{booktitle}{\emph{Proceedings of the 29th ACM SIGPLAN Conference on
  Programming Language Design and Implementation}} \emph{(\bibinfo{series}{PLDI
  '08})}.
\newblock


\bibitem[\protect\citeauthoryear{Godefroid, Klarlund, and Sen}{Godefroid
  et~al\mbox{.}}{2005}]%
        {Godefroid05}
\bibfield{author}{\bibinfo{person}{Patrice Godefroid}, \bibinfo{person}{Nils
  Klarlund}, {and} \bibinfo{person}{Koushik Sen}.}
  \bibinfo{year}{2005}\natexlab{}.
\newblock \showarticletitle{{DART: Directed Automated Random Testing}}. In
  \bibinfo{booktitle}{\emph{Proceedings of the 2005 ACM SIGPLAN Conference on
  Programming Language Design and Implementation}} \emph{(\bibinfo{series}{PLDI
  '05})}.
\newblock


\bibitem[\protect\citeauthoryear{Godefroid, Peleg, and Singh}{Godefroid
  et~al\mbox{.}}{2017}]%
        {Godefroid17}
\bibfield{author}{\bibinfo{person}{Patrice Godefroid}, \bibinfo{person}{Hila
  Peleg}, {and} \bibinfo{person}{Rishabh Singh}.}
  \bibinfo{year}{2017}\natexlab{}.
\newblock \showarticletitle{Learn \& Fuzz: Machine Learning for Input Fuzzing}.
  In \bibinfo{booktitle}{\emph{Proceedings of the 32Nd IEEE/ACM International
  Conference on Automated Software Engineering}} \emph{(\bibinfo{series}{ASE
  2017})}. \bibinfo{publisher}{IEEE Press}, \bibinfo{address}{Piscataway, NJ,
  USA}, \bibinfo{pages}{50--59}.
\newblock
\showISBNx{978-1-5386-2684-9}
\urldef\tempurl%
\url{http://dl.acm.org/citation.cfm?id=3155562.3155573}
\showURL{%
\tempurl}


\bibitem[\protect\citeauthoryear{Hoffmann, Janiczak, and Mandrikov}{Hoffmann
  et~al\mbox{.}}{2011}]%
        {jacoco}
\bibfield{author}{\bibinfo{person}{Marc~R Hoffmann}, \bibinfo{person}{B
  Janiczak}, {and} \bibinfo{person}{E Mandrikov}.}
  \bibinfo{year}{2011}\natexlab{}.
\newblock \bibinfo{title}{Eclemma-jacoco java code coverage library}.
\newblock
\newblock


\bibitem[\protect\citeauthoryear{Holler, Herzig, and Zeller}{Holler
  et~al\mbox{.}}{2012}]%
        {Holler12}
\bibfield{author}{\bibinfo{person}{Christian Holler}, \bibinfo{person}{Kim
  Herzig}, {and} \bibinfo{person}{Andreas Zeller}.}
  \bibinfo{year}{2012}\natexlab{}.
\newblock \showarticletitle{Fuzzing with Code Fragments}. In
  \bibinfo{booktitle}{\emph{Presented as part of the 21st {USENIX} Security
  Symposium ({USENIX} Security 12)}}.
\newblock


\bibitem[\protect\citeauthoryear{Holser}{Holser}{2014}]%
        {junit-quickcheck}
\bibfield{author}{\bibinfo{person}{Paul Holser}.}
  \bibinfo{year}{2014}\natexlab{}.
\newblock \bibinfo{title}{{junit-quickcheck: Property-based testing,
  JUnit-style}}.
\newblock
  \bibinfo{howpublished}{\url{https://pholser.github.io/junit-quickcheck}}.
\newblock
\newblock
\shownote{Accessed January 11, 2019.}


\bibitem[\protect\citeauthoryear{Infrastructure}{Infrastructure}{2016}]%
        {libFuzzer}
\bibfield{author}{\bibinfo{person}{LLVM~Compiler Infrastructure}.}
  \bibinfo{year}{2016}\natexlab{}.
\newblock \bibinfo{title}{libFuzzer}.
\newblock \bibinfo{howpublished}{\url{https://llvm.org/docs/LibFuzzer.html}}.
\newblock
\newblock
\shownote{Accessed April 17, 2019.}


\bibitem[\protect\citeauthoryear{King}{King}{1976}]%
        {King76}
\bibfield{author}{\bibinfo{person}{James~C. King}.}
  \bibinfo{year}{1976}\natexlab{}.
\newblock \showarticletitle{Symbolic execution and program testing}.
\newblock \bibinfo{journal}{\emph{Commun. ACM}}  \bibinfo{volume}{19}
  (\bibinfo{date}{July} \bibinfo{year}{1976}), \bibinfo{pages}{385--394}.
\newblock
Issue 7.
\showISSN{0001-0782}


\bibitem[\protect\citeauthoryear{Klees, Ruef, Cooper, Wei, and Hicks}{Klees
  et~al\mbox{.}}{2018}]%
        {Klees18}
\bibfield{author}{\bibinfo{person}{George Klees}, \bibinfo{person}{Andrew
  Ruef}, \bibinfo{person}{Benji Cooper}, \bibinfo{person}{Shiyi Wei}, {and}
  \bibinfo{person}{Michael Hicks}.} \bibinfo{year}{2018}\natexlab{}.
\newblock \showarticletitle{Evaluating Fuzz Testing}. In
  \bibinfo{booktitle}{\emph{Proceedings of the 2018 ACM SIGSAC Conference on
  Computer and Communications Security}} \emph{(\bibinfo{series}{CCS '18})}.
  \bibinfo{publisher}{ACM}, \bibinfo{address}{New York, NY, USA},
  \bibinfo{pages}{2123--2138}.
\newblock
\showISBNx{978-1-4503-5693-0}
\urldef\tempurl%
\url{https://doi.org/10.1145/3243734.3243804}
\showDOI{\tempurl}


\bibitem[\protect\citeauthoryear{Laeufer, Koenig, Kim, Bachrach, and
  Sen}{Laeufer et~al\mbox{.}}{2018}]%
        {Laeufer18}
\bibfield{author}{\bibinfo{person}{Kevin Laeufer}, \bibinfo{person}{Jack
  Koenig}, \bibinfo{person}{Donggyu Kim}, \bibinfo{person}{Jonathan Bachrach},
  {and} \bibinfo{person}{Koushik Sen}.} \bibinfo{year}{2018}\natexlab{}.
\newblock \showarticletitle{RFUZZ: Coverage-directed Fuzz Testing of RTL on
  FPGAs}. In \bibinfo{booktitle}{\emph{Proceedings of the International
  Conference on Computer-Aided Design}} \emph{(\bibinfo{series}{ICCAD '18})}.
  \bibinfo{publisher}{ACM}, \bibinfo{address}{New York, NY, USA}, Article
  \bibinfo{articleno}{28}, \bibinfo{numpages}{8}~pages.
\newblock
\showISBNx{978-1-4503-5950-4}
\urldef\tempurl%
\url{https://doi.org/10.1145/3240765.3240842}
\showDOI{\tempurl}


\bibitem[\protect\citeauthoryear{Lampropoulos and Sagonas}{Lampropoulos and
  Sagonas}{2012}]%
        {Lampropoulos12}
\bibfield{author}{\bibinfo{person}{Leonidas Lampropoulos} {and}
  \bibinfo{person}{Konstantinos Sagonas}.} \bibinfo{year}{2012}\natexlab{}.
\newblock \showarticletitle{Automatic WSDL-guided Test Case Generation for
  PropEr Testing of Web Services}. In \bibinfo{booktitle}{\emph{Proceedings 8th
  International Workshop on Automated Specification and Verification of Web
  Systems, {WWV} 2012, Stockholm, Sweden, 16th July 2012.}}
  \bibinfo{pages}{3--16}.
\newblock
\urldef\tempurl%
\url{https://doi.org/10.4204/EPTCS.98.3}
\showDOI{\tempurl}


\bibitem[\protect\citeauthoryear{Lemieux and Sen}{Lemieux and Sen}{2018}]%
        {Lemieux18-FairFuzz}
\bibfield{author}{\bibinfo{person}{Caroline Lemieux} {and}
  \bibinfo{person}{Koushik Sen}.} \bibinfo{year}{2018}\natexlab{}.
\newblock \showarticletitle{{FairFuzz: A Targeted Mutation Strategy for
  Increasing Greybox Fuzz Testing Coverage}}. In
  \bibinfo{booktitle}{\emph{Proceedings of the 33rd ACM/IEEE International
  Conference on Automated Software Engineering}} \emph{(\bibinfo{series}{ASE
  '18})}.
\newblock


\bibitem[\protect\citeauthoryear{Li, Ghosh, and Rajan}{Li
  et~al\mbox{.}}{2011}]%
        {Li11}
\bibfield{author}{\bibinfo{person}{Guodong Li}, \bibinfo{person}{Indradeep
  Ghosh}, {and} \bibinfo{person}{Sreeranga~P. Rajan}.}
  \bibinfo{year}{2011}\natexlab{}.
\newblock \showarticletitle{KLOVER: A Symbolic Execution and Automatic Test
  Generation Tool for C++ Programs}. In \bibinfo{booktitle}{\emph{CAV}}.
  \bibinfo{pages}{609--615}.
\newblock


\bibitem[\protect\citeauthoryear{Li, Chen, Chandramohan, Lin, Liu, and Tiu}{Li
  et~al\mbox{.}}{2017}]%
        {Li17}
\bibfield{author}{\bibinfo{person}{Yuekang Li}, \bibinfo{person}{Bihuan Chen},
  \bibinfo{person}{Mahinthan Chandramohan}, \bibinfo{person}{Shang-Wei Lin},
  \bibinfo{person}{Yang Liu}, {and} \bibinfo{person}{Alwen Tiu}.}
  \bibinfo{year}{2017}\natexlab{}.
\newblock \showarticletitle{Steelix: Program-state Based Binary Fuzzing}. In
  \bibinfo{booktitle}{\emph{Proceedings of the 2017 11th Joint Meeting on
  Foundations of Software Engineering}} \emph{(\bibinfo{series}{ESEC/FSE
  2017})}.
\newblock


\bibitem[\protect\citeauthoryear{Lindholm, Yellin, Bracha, and
  Buckley}{Lindholm et~al\mbox{.}}{2014}]%
        {jvm8-spec}
\bibfield{author}{\bibinfo{person}{Tim Lindholm}, \bibinfo{person}{Frank
  Yellin}, \bibinfo{person}{Gilad Bracha}, {and} \bibinfo{person}{Alex
  Buckley}.} \bibinfo{year}{2014}\natexlab{}.
\newblock \bibinfo{booktitle}{\emph{The Java Virtual Machine Specification,
  Java SE 8 Edition} (\bibinfo{edition}{1st} ed.)}.
\newblock \bibinfo{publisher}{Addison-Wesley Professional}.
\newblock
\showISBNx{013390590X, 9780133905908}


\bibitem[\protect\citeauthoryear{L\"{o}scher and Sagonas}{L\"{o}scher and
  Sagonas}{2017}]%
        {Loscher17}
\bibfield{author}{\bibinfo{person}{Andreas L\"{o}scher} {and}
  \bibinfo{person}{Konstantinos Sagonas}.} \bibinfo{year}{2017}\natexlab{}.
\newblock \showarticletitle{Targeted Property-based Testing}. In
  \bibinfo{booktitle}{\emph{Proceedings of the 26th ACM SIGSOFT International
  Symposium on Software Testing and Analysis}} \emph{(\bibinfo{series}{ISSTA
  2017})}. \bibinfo{publisher}{ACM}, \bibinfo{address}{New York, NY, USA},
  \bibinfo{pages}{46--56}.
\newblock
\showISBNx{978-1-4503-5076-1}
\urldef\tempurl%
\url{https://doi.org/10.1145/3092703.3092711}
\showDOI{\tempurl}


\bibitem[\protect\citeauthoryear{Loscher and Sagonas}{Loscher and
  Sagonas}{2018}]%
        {Loscher18}
\bibfield{author}{\bibinfo{person}{A. Loscher} {and} \bibinfo{person}{K.
  Sagonas}.} \bibinfo{year}{2018}\natexlab{}.
\newblock \showarticletitle{Automating Targeted Property-Based Testing}. In
  \bibinfo{booktitle}{\emph{2018 IEEE 11th International Conference on Software
  Testing, Verification and Validation (ICST)}}, Vol.~\bibinfo{volume}{00}.
  \bibinfo{pages}{70--80}.
\newblock
\urldef\tempurl%
\url{https://doi.org/10.1109/ICST.2018.00017}
\showDOI{\tempurl}


\bibitem[\protect\citeauthoryear{Maurer}{Maurer}{1990}]%
        {Maurer90}
\bibfield{author}{\bibinfo{person}{Peter~M. Maurer}.}
  \bibinfo{year}{1990}\natexlab{}.
\newblock \showarticletitle{Generating test data with enhanced context-free
  grammars}.
\newblock \bibinfo{journal}{\emph{Ieee Software}} \bibinfo{volume}{7},
  \bibinfo{number}{4} (\bibinfo{year}{1990}), \bibinfo{pages}{50--55}.
\newblock


\bibitem[\protect\citeauthoryear{Ognawala, Hutzelmann, Psallida, and
  Pretschner}{Ognawala et~al\mbox{.}}{2018}]%
        {Ognawala18}
\bibfield{author}{\bibinfo{person}{Saahil Ognawala}, \bibinfo{person}{Thomas
  Hutzelmann}, \bibinfo{person}{Eirini Psallida}, {and}
  \bibinfo{person}{Alexander Pretschner}.} \bibinfo{year}{2018}\natexlab{}.
\newblock \showarticletitle{Improving Function Coverage with Munch: A Hybrid
  Fuzzing and Directed Symbolic Execution Approach}. In
  \bibinfo{booktitle}{\emph{Proceedings of the 33rd Annual ACM Symposium on
  Applied Computing}} \emph{(\bibinfo{series}{SAC '18})}.
  \bibinfo{publisher}{ACM}, \bibinfo{address}{New York, NY, USA},
  \bibinfo{pages}{1475--1482}.
\newblock
\showISBNx{978-1-4503-5191-1}
\urldef\tempurl%
\url{https://doi.org/10.1145/3167132.3167289}
\showDOI{\tempurl}


\bibitem[\protect\citeauthoryear{{OW2 Consortium}}{{OW2 Consortium}}{2018}]%
        {asm}
\bibfield{author}{\bibinfo{person}{{OW2 Consortium}}.}
  \bibinfo{year}{2018}\natexlab{}.
\newblock \bibinfo{title}{{ObjectWeb ASM}}.
\newblock \bibinfo{howpublished}{\url{https://asm.ow2.io}}.
\newblock
\newblock
\shownote{Accessed August 21, 2018.}


\bibitem[\protect\citeauthoryear{Pacheco, Lahiri, Ernst, and Ball}{Pacheco
  et~al\mbox{.}}{2007}]%
        {Pacheco07}
\bibfield{author}{\bibinfo{person}{Carlos Pacheco},
  \bibinfo{person}{Shuvendu~K. Lahiri}, \bibinfo{person}{Michael~D. Ernst},
  {and} \bibinfo{person}{Thomas Ball}.} \bibinfo{year}{2007}\natexlab{}.
\newblock \showarticletitle{Feedback-Directed Random Test Generation}. In
  \bibinfo{booktitle}{\emph{Proceedings of the 29th International Conference on
  Software Engineering}} \emph{(\bibinfo{series}{ICSE '07})}.
  \bibinfo{publisher}{IEEE Computer Society}, \bibinfo{address}{Washington, DC,
  USA}, \bibinfo{pages}{75--84}.
\newblock
\showISBNx{0-7695-2828-7}
\urldef\tempurl%
\url{https://doi.org/10.1109/ICSE.2007.37}
\showDOI{\tempurl}


\bibitem[\protect\citeauthoryear{Padhye, Lemieux, and Sen}{Padhye
  et~al\mbox{.}}{2019}]%
        {Padhye19-tool}
\bibfield{author}{\bibinfo{person}{Rohan Padhye}, \bibinfo{person}{Caroline
  Lemieux}, {and} \bibinfo{person}{Koushik Sen}.}
  \bibinfo{year}{2019}\natexlab{}.
\newblock \showarticletitle{{JQF}: Coverage-guided Property-based Testing in
  {Java}}. In \bibinfo{booktitle}{\emph{Proceedings of the 28th ACM SIGSOFT
  International Symposium on Software Testing and Analysis}}
  \emph{(\bibinfo{series}{ISSTA '19})}.
\newblock
\urldef\tempurl%
\url{https://doi.org/10.1145/3293882.3339002}
\showDOI{\tempurl}


\bibitem[\protect\citeauthoryear{Papadakis and Sagonas}{Papadakis and
  Sagonas}{2011}]%
        {Papadakis11}
\bibfield{author}{\bibinfo{person}{Manolis Papadakis} {and}
  \bibinfo{person}{Konstantinos Sagonas}.} \bibinfo{year}{2011}\natexlab{}.
\newblock \showarticletitle{A PropEr Integration of Types and Function
  Specifications with Property-based Testing}. In
  \bibinfo{booktitle}{\emph{Proceedings of the 10th ACM SIGPLAN Workshop on
  Erlang}} \emph{(\bibinfo{series}{Erlang '11})}. \bibinfo{publisher}{ACM},
  \bibinfo{address}{New York, NY, USA}, \bibinfo{pages}{39--50}.
\newblock
\showISBNx{978-1-4503-0859-5}
\urldef\tempurl%
\url{https://doi.org/10.1145/2034654.2034663}
\showDOI{\tempurl}


\bibitem[\protect\citeauthoryear{Pham, B{\"{o}}hme, Santosa, Caciulescu, and
  Roychoudhury}{Pham et~al\mbox{.}}{2018}]%
        {Pham18}
\bibfield{author}{\bibinfo{person}{Van{-}Thuan Pham}, \bibinfo{person}{Marcel
  B{\"{o}}hme}, \bibinfo{person}{Andrew~E. Santosa},
  \bibinfo{person}{Alexandru~Razvan Caciulescu}, {and} \bibinfo{person}{Abhik
  Roychoudhury}.} \bibinfo{year}{2018}\natexlab{}.
\newblock \showarticletitle{Smart Greybox Fuzzing}.
\newblock \bibinfo{journal}{\emph{CoRR}}  \bibinfo{volume}{abs/1811.09447}
  (\bibinfo{year}{2018}).
\newblock
\showeprint[arxiv]{1811.09447}
\urldef\tempurl%
\url{http://arxiv.org/abs/1811.09447}
\showURL{%
\tempurl}


\bibitem[\protect\citeauthoryear{Rawat, Jain, Kumar, Cojocar, Giuffrida, and
  Bos}{Rawat et~al\mbox{.}}{2017}]%
        {Rawat17}
\bibfield{author}{\bibinfo{person}{Sanjay Rawat}, \bibinfo{person}{Vivek Jain},
  \bibinfo{person}{Ashish Kumar}, \bibinfo{person}{Lucian Cojocar},
  \bibinfo{person}{Cristiano Giuffrida}, {and} \bibinfo{person}{Herbert Bos}.}
  \bibinfo{year}{2017}\natexlab{}.
\newblock \showarticletitle{VUzzer: Application-aware Evolutionary Fuzzing}. In
  \bibinfo{booktitle}{\emph{Proceedings of the 2017 Network and Distributed
  System Security Symposium}} \emph{(\bibinfo{series}{NDSS '17})}.
\newblock


\bibitem[\protect\citeauthoryear{Rebert, Cha, Avgerinos, Foote, Warren, Grieco,
  and Brumley}{Rebert et~al\mbox{.}}{2014}]%
        {Rebert14}
\bibfield{author}{\bibinfo{person}{Alexandre Rebert}, \bibinfo{person}{Sang~Kil
  Cha}, \bibinfo{person}{Thanassis Avgerinos}, \bibinfo{person}{Jonathan
  Foote}, \bibinfo{person}{David Warren}, \bibinfo{person}{Gustavo Grieco},
  {and} \bibinfo{person}{David Brumley}.} \bibinfo{year}{2014}\natexlab{}.
\newblock \showarticletitle{Optimizing Seed Selection for Fuzzing}. In
  \bibinfo{booktitle}{\emph{Proceedings of the 23rd USENIX Conference on
  Security Symposium}} \emph{(\bibinfo{series}{SEC'14})}.
  \bibinfo{publisher}{USENIX Association}, \bibinfo{address}{Berkeley, CA,
  USA}, \bibinfo{pages}{861--875}.
\newblock
\showISBNx{978-1-931971-15-7}
\urldef\tempurl%
\url{http://dl.acm.org/citation.cfm?id=2671225.2671280}
\showURL{%
\tempurl}


\bibitem[\protect\citeauthoryear{Sen, Marinov, and Agha}{Sen
  et~al\mbox{.}}{2005}]%
        {Sen05}
\bibfield{author}{\bibinfo{person}{Koushik Sen}, \bibinfo{person}{Darko
  Marinov}, {and} \bibinfo{person}{Gul Agha}.} \bibinfo{year}{2005}\natexlab{}.
\newblock \showarticletitle{CUTE: A Concolic Unit Testing Engine for C}. In
  \bibinfo{booktitle}{\emph{Proceedings of the 10th European Software
  Engineering Conference Held Jointly with 13th ACM SIGSOFT International
  Symposium on Foundations of Software Engineering}}
  \emph{(\bibinfo{series}{ESEC/FSE-13})}.
\newblock


\bibitem[\protect\citeauthoryear{Serebryany, Buka, and Morehouse}{Serebryany
  et~al\mbox{.}}{2017}]%
        {Serebryany17}
\bibfield{author}{\bibinfo{person}{Kostya Serebryany}, \bibinfo{person}{Vitaly
  Buka}, {and} \bibinfo{person}{Matt Morehouse}.}
  \bibinfo{year}{2017}\natexlab{}.
\newblock \bibinfo{title}{Structure-aware fuzzing for {Clang} and {LLVM} with
  libprotobuf-mutator}.
\newblock
\newblock


\bibitem[\protect\citeauthoryear{Sirer and Bershad}{Sirer and Bershad}{1999}]%
        {Sirer99}
\bibfield{author}{\bibinfo{person}{Emin~G\"{u}n Sirer} {and}
  \bibinfo{person}{Brian~N. Bershad}.} \bibinfo{year}{1999}\natexlab{}.
\newblock \showarticletitle{Using Production Grammars in Software Testing}. In
  \bibinfo{booktitle}{\emph{Proceedings of the 2Nd Conference on
  Domain-specific Languages}} \emph{(\bibinfo{series}{DSL '99})}.
  \bibinfo{publisher}{ACM}, \bibinfo{address}{New York, NY, USA},
  \bibinfo{pages}{1--13}.
\newblock
\showISBNx{1-58113-255-7}
\urldef\tempurl%
\url{https://doi.org/10.1145/331960.331965}
\showDOI{\tempurl}


\bibitem[\protect\citeauthoryear{Stephens, Grosen, Salls, Dutcher, Wang,
  Corbetta, Shoshitaishvili, Kruegel, and Vigna}{Stephens
  et~al\mbox{.}}{2016}]%
        {Stephens16}
\bibfield{author}{\bibinfo{person}{Nick Stephens}, \bibinfo{person}{John
  Grosen}, \bibinfo{person}{Christopher Salls}, \bibinfo{person}{Andrew
  Dutcher}, \bibinfo{person}{Ruoyu Wang}, \bibinfo{person}{Jacopo Corbetta},
  \bibinfo{person}{Yan Shoshitaishvili}, \bibinfo{person}{Christopher Kruegel},
  {and} \bibinfo{person}{Giovanni Vigna}.} \bibinfo{year}{2016}\natexlab{}.
\newblock \showarticletitle{Driller: Augmenting Fuzzing Through Selective
  Symbolic Execution}. In \bibinfo{booktitle}{\emph{Proceedings of the 2016
  Network and Distributed System Security Symposium}}
  \emph{(\bibinfo{series}{NDSS '16})}.
\newblock


\bibitem[\protect\citeauthoryear{Tillmann and de~Halleux}{Tillmann and
  de~Halleux}{2008}]%
        {Tillmann08}
\bibfield{author}{\bibinfo{person}{Nikolai Tillmann} {and}
  \bibinfo{person}{Jonathan de Halleux}.} \bibinfo{year}{2008}\natexlab{}.
\newblock \showarticletitle{Pex - White Box Test Generation for {{.NET}}}. In
  \bibinfo{booktitle}{\emph{Proceedings of Tests and Proofs}}.
\newblock


\bibitem[\protect\citeauthoryear{Wang, Chen, Wei, and Liu}{Wang
  et~al\mbox{.}}{2017}]%
        {Wang17}
\bibfield{author}{\bibinfo{person}{J. Wang}, \bibinfo{person}{B. Chen},
  \bibinfo{person}{L. Wei}, {and} \bibinfo{person}{Y. Liu}.}
  \bibinfo{year}{2017}\natexlab{}.
\newblock \showarticletitle{Skyfire: Data-Driven Seed Generation for Fuzzing}.
  In \bibinfo{booktitle}{\emph{2017 IEEE Symposium on Security and Privacy
  (SP)}} \emph{(\bibinfo{series}{SP '17})}.
\newblock


\bibitem[\protect\citeauthoryear{Wang, Chen, Wei, and Liu}{Wang
  et~al\mbox{.}}{2019}]%
        {Wang19}
\bibfield{author}{\bibinfo{person}{Junjie Wang}, \bibinfo{person}{Bihuan Chen},
  \bibinfo{person}{Lei Wei}, {and} \bibinfo{person}{Yang Liu}.}
  \bibinfo{year}{2019}\natexlab{}.
\newblock \showarticletitle{Superion: Grammar-Aware Greybox Fuzzing}. In
  \bibinfo{booktitle}{\emph{41st International Conference on Software
  Engineering}} \emph{(\bibinfo{series}{ICSE '19})}.
\newblock


\bibitem[\protect\citeauthoryear{Yang, Chen, Eide, and Regehr}{Yang
  et~al\mbox{.}}{2011}]%
        {Yang11}
\bibfield{author}{\bibinfo{person}{Xuejun Yang}, \bibinfo{person}{Yang Chen},
  \bibinfo{person}{Eric Eide}, {and} \bibinfo{person}{John Regehr}.}
  \bibinfo{year}{2011}\natexlab{}.
\newblock \showarticletitle{{Finding and Understanding Bugs in C Compilers}}.
  In \bibinfo{booktitle}{\emph{Proceedings of the 32nd ACM SIGPLAN Conference
  on Programming Language Design and Implementation}}
  \emph{(\bibinfo{series}{PLDI '11})}.
\newblock


\bibitem[\protect\citeauthoryear{Yun, Lee, Xu, Jang, and Kim}{Yun
  et~al\mbox{.}}{2018}]%
        {Yun18}
\bibfield{author}{\bibinfo{person}{Insu Yun}, \bibinfo{person}{Sangho Lee},
  \bibinfo{person}{Meng Xu}, \bibinfo{person}{Yeongjin Jang}, {and}
  \bibinfo{person}{Taesoo Kim}.} \bibinfo{year}{2018}\natexlab{}.
\newblock \showarticletitle{QSYM: A Practical Concolic Execution Engine
  Tailored for Hybrid Fuzzing}. In \bibinfo{booktitle}{\emph{Proceedings of the
  27th USENIX Conference on Security Symposium}}
  \emph{(\bibinfo{series}{SEC'18})}. \bibinfo{publisher}{USENIX Association},
  \bibinfo{address}{Berkeley, CA, USA}, \bibinfo{pages}{745--761}.
\newblock
\showISBNx{978-1-931971-46-1}
\urldef\tempurl%
\url{http://dl.acm.org/citation.cfm?id=3277203.3277260}
\showURL{%
\tempurl}


\bibitem[\protect\citeauthoryear{Zalewski}{Zalewski}{2014}]%
        {afl}
\bibfield{author}{\bibinfo{person}{Micha\l{} Zalewski}.}
  \bibinfo{year}{2014}\natexlab{}.
\newblock \bibinfo{title}{American Fuzzy Lop}.
\newblock \bibinfo{howpublished}{\url{http://lcamtuf.coredump.cx/afl}}.
\newblock
\newblock
\shownote{Accessed January 11, 2019.}


\bibitem[\protect\citeauthoryear{Zalewski}{Zalewski}{2016}]%
        {FidgetyAFL}
\bibfield{author}{\bibinfo{person}{Micha\l{} Zalewski}.}
  \bibinfo{year}{2016}\natexlab{}.
\newblock \bibinfo{title}{{FidgetyAFL}}.
\newblock
  \bibinfo{howpublished}{\url{https://groups.google.com/d/msg/afl-users/fOPeb62FZUg/CES5lhznDgAJ}}.
\newblock
\newblock
\shownote{Accessed Jan 28th, 2019.}


\end{thebibliography}

\end{document}